\documentclass[fleqn,usenatbib]{mnras}

\usepackage{newtxtext,newtxmath}

\usepackage[T1]{fontenc}

\DeclareRobustCommand{\VAN}[3]{#2}
\let\VANthebibliography\thebibliography
\def\thebibliography{\DeclareRobustCommand{\VAN}[3]{##3}\VANthebibliography}

\usepackage{graphicx}	
\usepackage{multirow}   
\usepackage{hyperref}

\title[A Multi-Bands Catalogue on the North Ecliptic Pole]{Identification of AKARI infrared sources by Deep HSC Optical Survey: \\
Construction of New Band-Merged Catalogue in the NEP-Wide field}

\author[S. J. Kim et al.]{
Seong Jin Kim,$^{1}$\thanks{E-mail: seongini@phys.nthu.edu.tw; seongini@gmail.com}
Nagisa Oi,$^{2}$
Tomotsugu Goto,$^{1}$
Hiroyuki Ikeda,$^{3,4}$
Simon C.-C. Ho,$^{1}$
Hyunjin Shim,$^{5}$
\newauthor
Yoshiki Toba,$^{6,7,8}$
Ho Seong Hwang,$^{9}$
Tetsuya Hashimoto,$^{1,10}$
Laia Barrufet,$^{11,12}$
Matthew Malkan,$^{13}$
\newauthor
Helen K. Kim,$^{13}$
Ting-Chi Huang,$^{14,15}$
Hideo Matsuhara,$^{15}$
Takamitsu Miyaji,$^{16,17,*}$
Chris Pearson,$^{12,18,19}$
\newauthor
Stephen Serjeant,$^{18}$
Daryl Joe Santos,$^{1}$
Eunbin Kim,$^{9}$
Agnieszka  Pollo,$^{20,21}$
Woong-Seob Jeong,$^{9}$
\newauthor
Ting-Wen Wang,$^{1}$
Rieko Momose,$^{22}$
and Toshinobu Takagi$^{14}$
\\
$^{1}$Institute of Astronomy, National Tsing Hua University, 101, Section 2. Kuang-Fu Road, Hsinchu, 30013, Taiwan\\
$^{2}$Tokyo University of Science, 1-3, Kagurazaka Shinjuku Tokyo 162-8601 Japan\\
$^{3}$National Astronomical Observatory of Japan, 2-21-1 Osawa, Mitaka, Tokyo 181-8588, Japan\\
$^{4}$National Institute of Technology, Wakayama College, 77 Noshima, Nada-cho, Gobo, Wakayama 644-0023, Japan\\
$^{5}$Department of Earth Science Education, Kyungpook National University, Daegu 41566, Korea\\
$^{6}$Department of Astronomy, Kyoto University,
Kitashirakawa-Oiwake-cho, Sakyo-ku, Kyoto 606-8502, Japan\\
$^{7}$Academia Sinica Institute of Astronomy and Astrophysics, 11F of Astronomy-Mathematics Building, AS/NTU, No.1, Section 4, Roosevelt Road, Taipei 10617, Taiwan\\
$^{8}$Research Center for Space and Cosmic Evolution, Ehime
University, 2-5 Bunkyo-cho, Matsuyama, Ehime 790-8577, Japan\\
$^{9}$Korea Astronomy and Space Science Institute (KASI), 776 Daedeok-daero, Yuseong-gu, Daejeon 34055, Korea\\
$^{10}$Centre for Informatics and Computation in Astronomy (CICA), National Tsing Hua University, 101, Section 2. Kuang-Fu Road, Hsinchu, 30013, Taiwan (R.O.C.)\\
$^{11}$European Space Astronoy Center, 28691 Villanueva de la Can\~ada, Spain\\
$^{12}$RAL Space, Rutherford Appleton Laboratory, Chilton, Didcot, Oxfordshire OX11 0QX, UK\\
$^{13}$Department of physics and astronomy, UCLA, 475 Portola Plaza, L.A., CA 90095-1547, USA\\
$^{14}$Department of Space and Astronautical Science, Graduate University for Advanced Studies, SOKENDAI, 
Hayama, Miura District, Kanagawa 240-0193, Japan\\
$^{15}$Institute of Space and Astronautical Science, Japan Aerospace Exploration Agency, Sagamihara, 252-5210 Kanagawa, Japan\\
$^{*}$On sabbatical leave from UNAM at AIP.\\
$^{16}$Instituto de Astronom\'ia, Universidad Nacional Aut\'onoma de M\'exico (UNAM), Km. 107, Carret. Tij.-Ens., Ensenada, 22860 Mexico\\
$^{17}$Leibniz Institut f\"ur Astrophysik Poysdam (AIP), An der Sternwarte, Potsdam, 14482, Germany\\
$^{18}$The Open University, Milton Keynes, MK7 6AA, UK\\
$^{19}$Oxford Astrophysics, University of Oxford, Keble Rd, Oxford OX1 3RH, UK\\
$^{20}$National Centre for Nuclear Research, ul. Pasteura 7, 02-093 Warsaw, Poland\\
$^{21}$Astronomical Observatory of the Jagiellonian University, ul. Orla 171, 30-244 Cracow, Poland\\
$^{22}$Department of Astronomy, School of Science, The University of Tokyo, 7-3-1 Hongo, Bunkyo-ky, Tokyo 113-0033, Japan\\
}

\date{Accepted XXX. Received YYY; in original form ZZZ}

\pubyear{2020}

\begin{document}
\label{firstpage}
\pagerange{\pageref{firstpage}--\pageref{lastpage}}
\maketitle

\begin{abstract}
The north ecliptic pole (NEP) field is a natural deep field location for many satellite observations. It has been targeted many times since it was surveyed by the AKARI space telescope with its unique wavelength coverage from the near- to mid-infrared (mid-IR). Many follow-up observations have been carried out and made this field one of the most frequently observed areas with a variety of facilities, accumulating abundant panchromatic data from X-ray to radio wavelength range. Recently, a deep optical survey with the Hyper Suprime-Cam (HSC) at the Subaru telescope covered the NEP-Wide (NEPW) field, which enabled us to identify faint sources in the near- and mid-IR bands, and to improve the photometric redshift (photo-z) estimation. In this work, we present newly identified AKARI sources by the HSC survey, along with multi-band  photometry for   91,861 AKARI sources observed over the NEPW field.  We release a new band-merged catalogue combining various photometric data from GALEX UV to the submillimetre (sub-mm) bands (e.g., Herschel/SPIRE, JCMT/SCUBA-2). About $\sim$ 20,000 AKARI sources are newly matched to the HSC data,  most of which seem to be faint galaxies in the near- to mid-infrared AKARI bands.
This catalogue motivates a variety of current research, and will be increasingly useful as recently launched (eROSITA/ART-XC) and future space missions (such as $JWST$, Euclid, and SPHEREx) plan to take deep observations in the NEP field.
\end{abstract}




\section{Introduction}  
To answer the questions concerning cosmic star formation history and galaxy evolution, it is critical to have a comprehensive understanding of the infrared (IR) luminous population of galaxies \citep{Casey14,Kirkpatrick12,Madau14,Sanders14}. 
They are presumably star-forming systems containing a large amount of dust, where the critical phase in their activities of star formation (SF) or active galactic nuclei (AGN) take place, hidden behind the obscuring dust \citep{Galliano18, Goto10, Hickox18, Lutz14}.

Wide-field cosmological surveys at IR wavelengths are the most efficient way to collect data for various populations of galaxies, especially for dusty star-forming galaxies (dusty SFGs, DSFGs) and obscured AGNs, at different cosmological epochs \citep{Matsuhara06,HHwang07,HHwang10,Toba15}.
Statistically significant samples of dusty galaxies based on large-area surveys covering significant cosmological volumes have to be obtained.
Also, follow-up surveys should be made to sample spectral energy distributions (SEDs): a comprehensive physical description requires wide wavelength coverage to capture the range of processes involved.
Most importantly, a deep optical follow-up survey is necessary because the optical identification is an essential prerequisite to understand nature of the sources \citep{Sutherland92, HHwang12},  e.g., star-galaxy separation or to derive photometric redshift (photo-z).

\begin{figure*}
\begin{center}
\resizebox{0.99\textwidth}{!}{\includegraphics{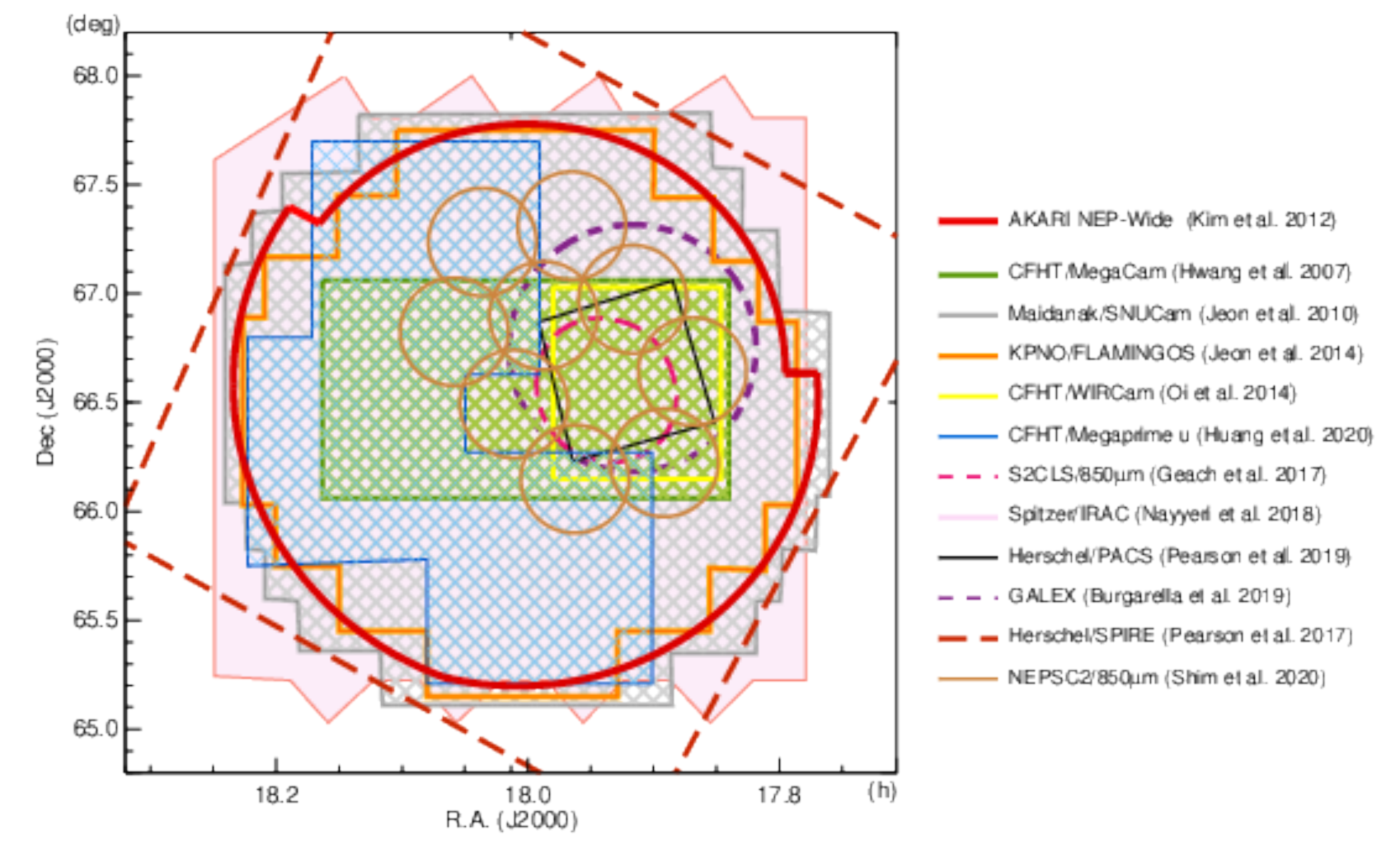}}
\caption{An overall map showing a variety of surveys around the NEP. The red circular area shows the AKARI's NEP-Wide field \citep{K12}.  The green and grey (meshed) areas represent the optical surveys done with the CFHT MegaCam \citep{H07}  and the Maidanak SNUCam \citep{J10}, respectively.  The yellow square shows a slightly deeper observation with the MegaCam as well as the WIRCam on the NEP-Deep field \citep{Oi14}. The area surrounded by blue line shows the additional u-band observation by the MegaPrime \citep{Huang20}. The pink shaded area indicates the recent near-IR survey with Spitzer \citep{Nayyeri18}. The small black square inside the yellow box shows the area observed by the Herschel/PACS \citep{Pearson19}. A broken magenta circle overlaid with the black square indicates the area observed by S2CLS \citep{Geach17}. Nine brown circles around the S2CLS  show the areas surveyed by NEPSC2 850 $\mu$m mapping program with SCUBA-2 \citep{Shim20}.
The largest rhombus (brown long-dashed line) shows the Herschel/SPIRE coverage \citep{Pearson17}. 
\label{fig01}}
\end{center}
\end{figure*}

The north ecliptic pole (NEP; $\alpha = 18^{h}00^{m}00^{s}$, $\delta = 66^{\circ}33^{\prime}88^{\prime\prime}$) has been a good target for deep, unbiased, and contiguous surveys for extra-galactic objects such as galaxies, galaxy clusters and AGNs because the NEP is a natural deep field location for a wide class of observatory missions \citep{Serjeant12}.  Many astronomical satellites, such as  ROSAT \citep{Henry06}, GALEX\footnote{http://www.galex.caltech.edu/} \citep{Burgarella19}, \textit{Spitzer}\footnote{http://www.spitzer.caltech.edu/} Space Telescope \citep{Jarrett11,Nayyeri18},  have accumulated a large number of exposures towards the NEP area because the Earth-orbiting satellites pass over the ecliptic poles and, for the Earth-trailing satellites, these poles are always in the continuous viewing zone. 
The AKARI \citep{Murakami07} also devoted a large amount of observing time (a span of more than a year) to cover a wide area over the NEP  using the infrared camera \citep[IRC,][]{Onaka07} with excellent visibility thanks to its polar sun-synchronous orbit \citep{Matsuhara06}.

A noticeable aspect of this NEP field surveyed by AKARI is that the ancillary optical data sets are abundant, supporting the identification of the infrared sources which enabled subsequent analyses properly \citep{K12}. In addition, many other surveys or follow-up observations on the NEP (but on a limited area) have been carried out from X-ray to radio wavelengths to cover the NEP area \citep{Krumpe15,Burgarella19,Pearson19,Geach17,White10,White17}  since the AKARI obtained valuable data sets (see Figure 1).
However, a fraction ($\sim 30\%$ at $N4$) of the IR sources detected by AKARI has been left unidentified by optical data because of the insufficient depths and incomplete areal coverage of the previous optical surveys. The different photometric filter systems used at different surveys also hampered homogeneous analyses based on unbiased sample selection, therefore a deeper and coherent optical survey on this field was necessarily required.


A new deep optical survey consistently covering the entire NEP-Wide (NEPW) field was carried out by the Hyper Suprime-Cam \citep[HSC:][]{Miyazaki18}  with five photometric filter bands ($g$, $r$, $i$, $z$ and $y$). These HSC data were reduced  \citep{Oi20} by the recent version of the pipeline  \citep[\texttt{v6.5.3},][]{Bosch18}, which allowed the depth of the new optical data to reach down to $\sim$ 2 mag deeper (at $g$ and $i$ band) than the previous optical survey with Canada-France-Hawaii Telescope  \citep[CFHT,][]{H07}. 
In addition, supplemental observation using the CFHT/MegaPrime \citep{Huang20}  replenished the insufficient coverage of $u^*$-band from the previous CFHT surveys, which brings photo-z accuracy improvement along with this new HSC data \citep{Ho20}. The source matching and band merging process (see Section 2 for the details) have been encouraging various subsequent works such as the recent luminosity function (LF) update \citep{Goto19}, properties of mid-IR (MIR) galaxies detected at 250$\mu$m \citep{Kim19}, estimation of number fraction of AGN populations \citep{Chiang19}, study on high-z population \citep{Barrufet20},  obscured AGN activity \citep{Wang20},  merger fraction depending on star formation mode (Kim et al. in prep), AGN activities depending on the environments (Santos et al. in prep), machine learning algorithms to classify/separate IR sources (Poliszczuk et al. in  prep; Chen et al. in prep), cluster candidates finding (Huang et al. in prep), and even on the AKARI  sources without any HSC counterpart \citep{Toba20}. 
The science on the NEP initiated by AKARI is now entering a new era with a  momentum driven by Subaru/HSC observations as well as current survey projects, such as homogeneous spectroscopic survey (MMT2020A/B, PI: H. S. Hwang) and the 850 $\mu$m mapping over the entire NEP area using with the Submillimetre Common-User Bollometric Array 2 (SCUBA-2) at the James Clerk Maxwell Telescope  \citep{Shim20}. 
More extensive imaging observations with HSC are still on going  with spectroscopy with Keck/DEIMOS+MOSFIRE as part of the Hawaii Two-O project (H20)\footnote{https://project.ifa.hawaii.edu/h20}. Spitzer also finished its ultra deep NIR observations recently as one of the Legacy Surveys (PI: Capak) 
before it retired early 2020 to carry out precursor survey for Euclid \citep{Laureijs11}, the {\it James Webb Space Telescope} \citep[{\it JWST},][]{Gardner06}   and the Wide Field InfraRed Survey Telescope \citep[WFIRST:][]{Spergel15} over this field.  
The Spektr-RG was launched in 2019 to the L2 point of the Sun-Earth system (1.5 million km away from us) and eROSITA\citep{Merloni12} started mission towards the NEP. 
Spectro-Photometer for the History of the Universe, Epoch of Reinonization, and Ice Explorer \citep[SPHEREx:][]{Dore16,Dore18}  are also planning to target this field. 

The main goal of this work is to identify optical counterparts of the AKARI/NEPW sources with more reliable optical photometry of the HSC images  (even for the faint NIR sources), and cross-check with all available supplementary data covering this field to build the panchromatic data. 
We briefly describe various data supporting AKARI/NEPW data, but mostly focus on explaining how we matched sources and combined all the data together.
This paper is organised as follows.  Sec. 2 introduces the HSC and AKARI data, and gives the detailed process how we cross-matched the sources between them. In Sec. 3, we present the complementary data sets used to construct the multi bands catalogue.  We describe their optical-IR properties (in statistical ways) in Sec. 4.  Sec. 5 gives the summary and conclusions. All magnitudes are presented in AB the magnitude system.

\section{ Identification of the AKARI's NEP-Wide sources using deep HSC data  }

\subsection{AKARI NEP-Wide survey data}
The NEP-Wide (NEPW) survey \citep{Matsuhara06, Lee09, K12}, as one of the large area survey missions  
of the AKARI space telescope \citep{Murakami07}, has provided us with a unique IR data set, sampling the near- to mid-IR wavelength range without large gaps between the filter bands (the circular area surrounded by a thick red line in Figure 1).  In this program, they observed the 5.4 deg$^2$ circular area centred at the NEP using nine photometric bands covering the range from 2 to 25 $\mu$m continuously. The overall strategy of the survey was explained by \cite{Lee09}. \cite{K12} presented the description of the data reduction, source detection, photometry, and catalogue.  They also combined the nine separated catalogues  (i.e., for  $N2$, $N3$, $N4$ in the NIR, $S7$, $S9W$, $S11$ in the MIR-S, and $L15$, $L18W$, $L24$ in the MIR-L channel) along with the optical identification/photometry.  Before combining, they carefully discriminated the spurious objects and false detection, in order to confirm the validity of the IR sources: they tried to identify the optical counterparts using the CFHT \citep{H07} and Maidanak \citep{J10} data and then cross-checked against the NIR $J$, $K$ band data obtained from the KPNO/FLAMINGOS \citep{Jeon14}.

The number of sources detected at the nine IRC bands (\texttt{DETECT\_THRESH=3}, 
\texttt{DETECT\_MINAREA=5})  by 
\texttt{SExtractor} \citep{Bertin96}
are 87858, 104170, 96159, 15390, 18772, 15680, 13148, 15154, and 4019 (the detection limits are 21 mag in NIR, 19.5 - 19 mags in MIR-S, and 18.6 - 17.8 mags in MIR-L), respectively \citep{K12}. A significant fraction of these sources (17 \% of the $N2$, 26 \% of the $N3$, and 30 \% of $N4$ sources) did not have optical data (mostly because they are not detected in optical surveys). In addition, $\sim$ 4 \% of the NIR sources were finally rejected because they are detected at only one NIR band (e.g., $N2$, $N3$, or $N4$) and are strongly suspected to be false objects: they suffered from the `multiplexer bleeding trails' (MUX-bleeding) due to the characteristics of the InSb detector array \citep{Holloway86,Offenberg01}  so that the source detection near the MUX-bleeding was strongly affected by artificial effects and spurious objects near the abnormally bright pixels.  Also, the false detection caused by cosmic ray hits were serious at the $N4$ band mostly because the telescope time to take dithering frames was sometimes assigned to the telescope maneuvering (by the IRC03 observation template).  If a certain source were detected at only one NIR band, it could potentially be an artifact or false detection. Therefore, the sources detected at only one NIR band were excluded in the first release of the band-merged NEPW catalogue. In the MIR bands,  cosmic ray hits and other artifacts are not numerous, and so the sources detected by only one MIR band were included. 

Note that the AKARI  NEP-Deep (NEPD) survey data \citep{Wada08,Takagi12}, which is similar to the NEPW survey (with the consistent photometries and the same astrometric accuracy) but different in terms of the coverage (0.7 deg$^2$) and the depth ($\sim$ 0.5 mag deeper), is not included in this work.

We expect that the new optical data obtained by the HSC \citep{Oi20}  will allow us to identify more IR sources, most of which are faint in the IR as well as the optical bands. Also, we may be able to examine if there are any real NIR sources that have been rejected just because they did not have any counterpart against the other bands (from the optical to MIR). We, therefore, repeated the merging of nine AKARI bands without any exclusion process, in order to attempt to recover possibly real AKARI sources not included in the study of \cite{K12}.
The sources detected at least one AKARI band can be included in the new AKARI 9 bands merged catalogue. Spurious objects or artifacts can be excluded later if we find any, during further analyses.

 When we carried out this procedure, we began with the matching between $N2$ and $N3$ band first. After that, we used these results against the $N4$ band using $N3$ coordinates. In the case without $N3$ coordinates (i.e., a source detected at $N2$ but not detected at $N3$), we took $N2$ coordinates for the matching against $N4$. This process went through down to the $L18W$ or $L24$.  In the resulting catalogue, we kept the coordinates from the shortest and the longest bands in this matching process. Therefore,  if a certain source was detected at neither $N2$ nor $L24$ but at the other bands, then the coordinates information for the shortest band is from $N3$ and the longest from $L18W$.  
There is no systematic offset among the astrometry from different AKARI bands (all of them are fixed at the same focal plane).  We eventually registered  130,150 IR sources  in the new  NEPW catalogue, which were detected at least one of the AKARI bands from $N2$ to $L24$. 

\begin{figure}
\begin{center}
\resizebox{\columnwidth}{!}{\includegraphics{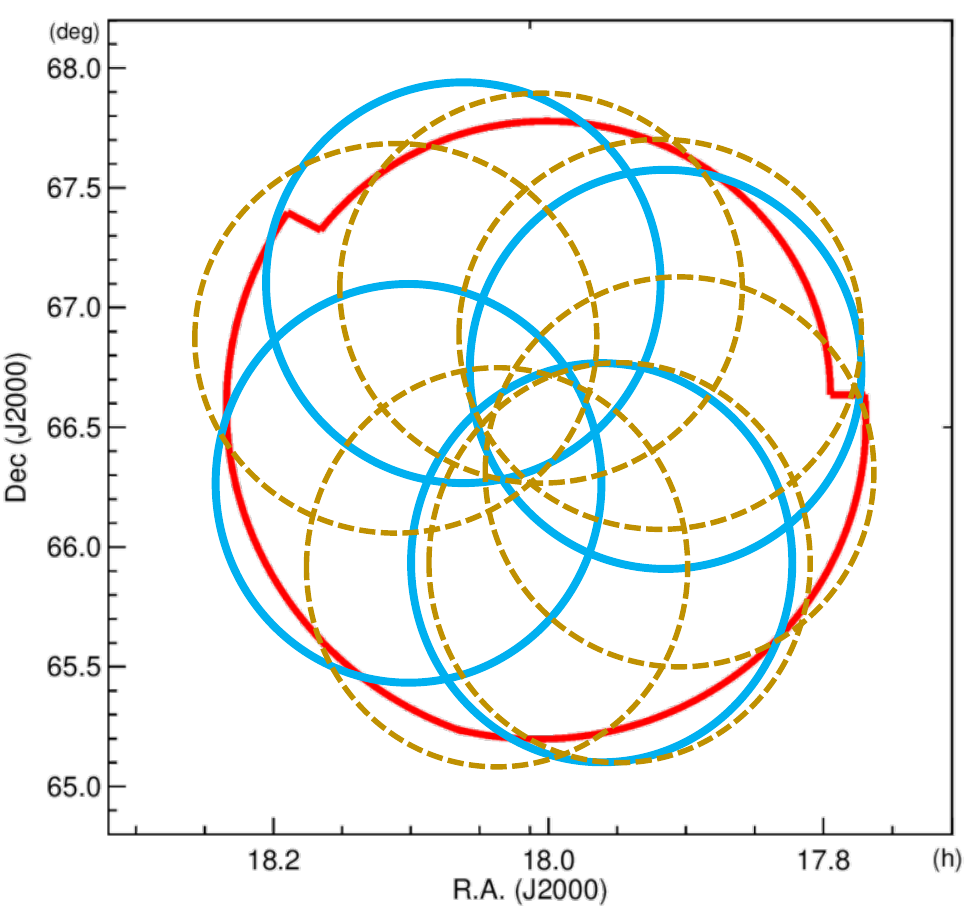}}
\caption{The map showing the HSC coverage over the NEPW survey area (a red circular area; Kim et al. 2012). The areas marked by yellow dashed circles show the region observed by the HSC $r$-band, while the four blue solid circles indicate the region observed by the $g$, $i$, $z$, and $y$-band (Oi et al. 2020). 
\label{fig02}}
\end{center}
\end{figure}

\subsection{Deep Optical Survey with HSC over the NEP}

A deep optical survey covering the whole NEP field with the HSC was proposed \citep{Goto17} in order to detect all the IR sources observed by AKARI.  Two optical datasets, one obtained  with the CFHT   \citep[the central green square in Figure 1;][]{H07} and  one with Maidanak telescope \citep[the grey area in Figure 1;][]{J10} have been supporting the AKARI IR data. However, the depths (5$\sigma$) of these optical surveys (25.4 and 23.1 AB mag at $r^{\prime}$ and $R$ band, respectively) were insufficient to identify all the AKARI IR sources. A slightly deeper observation by MegaCam, \citep{Oi14} on a smaller field was also carried out, but the areal coverage (0.67 deg$^2$) was only for the NEPD field (a yellow box in Figure 1).

\cite{Goto17} intended to use the large field of view (FoV; 1.5 deg in diameter, see Figure 2) of the HSC so that the entire NEPW field of AKARI was able to be covered by taking only 4 FoVs (for the $g$, $i$, $z$, and $y$ bands; the blue circles in Figure 2). Ten FoVs, in total, were allocated in six nights \citep{Goto17, Oi20}  including $r$ band  to take the whole NEPW field using those five HSC filter bands. 
The $r$ band imaging was taken earlier during the first observation in 2014, where the observations suffered from air disturbance (including the dome shutter opening error), which made the seeing at $r$ band worse (1$^{\prime\prime}$.25) than those of the other four (0.7 -- 0.8$^{\prime\prime}$ ) obtained later in the second observations (Aug. 2015) \citep{Oi20}.

The data reduction was carried out with the official HSC pipeline, \texttt{hscPipe} 6.5.3 \citep{Bosch18}.  
Apart from the fundamental pre-processing procedures (e.g., bias, dark, flat, fringe, etc.), the performance in the sky subtraction and artifact rejection was enhanced in this recent version. In particular, the peak culling procedure was included to cull  the spurious or unreliable detections, which improved the source detection results in the pipeline process. Owing to the bad seeing ($\sim 1^{\prime\prime}.25 $) in the $r$ band, the source detection was carried out on the $gizy$ band stacked image. However the photometry was forced to be performed at all five bands. All these  procedures with recent pipeline (e.g., updated flag information of the \texttt{hscPipe} 6.5.3 \citep{Bosch18}) eventually helped resolve the issues that have been reported for a few years regarding false detections (e.g., damaged sources near the boundary or along the image edge of each frame, and effects from the saturation of bright stars, etc.).  The 5$\sigma$ detection limits are 28.6, 27.3, 26.7, 26.0, and 25.6 mag at $g$, $r$, $i$, $z$, and $y$, respectively.  The limiting magnitudes of the $g$, $r$, $i$, and $z$ band of the previous CFHT data were 26.1, 25.6, 24.8, and 24.0 mag, respectively. We, therefore, obtained  1.7 -- 2.5 mags deeper optical data at the corresponding filter bands (even though the effective wavelengths of the filters are slightly different; see Table 1). 
Finally, 
we catalogued 3.25 million sources observed by the optical survey with the HSC along with a large number of parameters appended from the HSC data pipeline.

The magnigude at each band was given in terms of the \texttt{Cmodel} photometry, which performs well for galaxies and asymptotically approaches PSF photometry for compact sources. The colours estimated with  \texttt{Cmodel flux\footnote{AB mag $ = -2.5 $ log$_{10}$(\texttt{Cmodel flux}) + 27}}  are robust against seeing conditions \citep{Huang18}.  As we mentioned in the previous paragraph, seeing conditions  for  $r$-band  are different from the other four bands, therefore by taking \texttt{Cmodel flux} to calculate colours, these different seeing condition effects can be compensated.


\begin{figure*}
\begin{center}
\resizebox{0.8\textwidth}{!}{\includegraphics{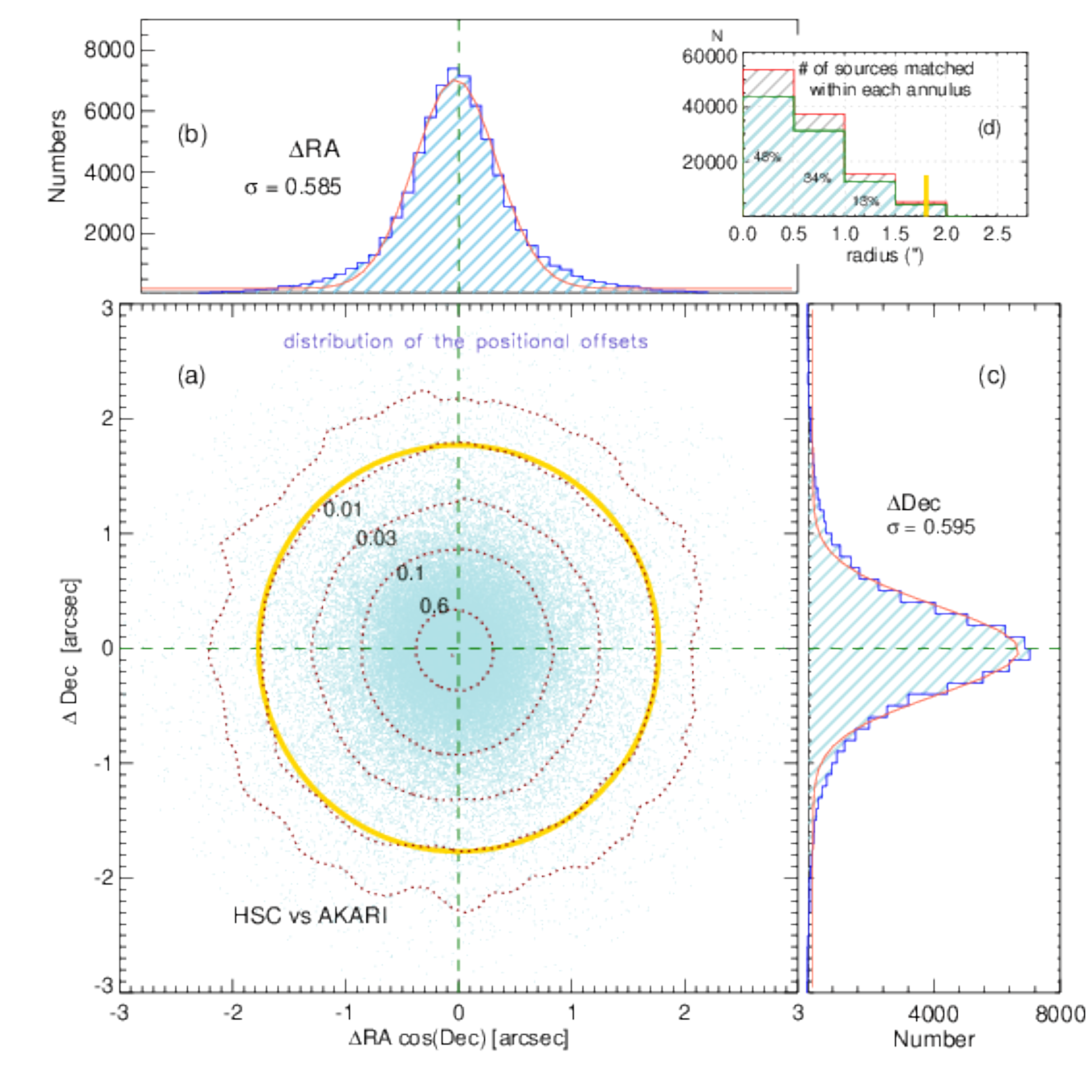}}
\caption{The positional offset distribution of the matched sources  between the HSC and AKARI data.  On the bottom left panel (a), the dotted contours with numbers represent the density levels normalised by the central peak density. The yellow circle is the matching radius determined based on the 3-sigma ($\sigma$) of the Gaussian, fitted to the histograms (magenta curves), on the top left (b) and bottom right (c) panels, shown in terms of 0.2$^{\prime \prime}$ bin. On a top right box (d), bars show how many sources were matched within the 0.5$^{\prime \prime}$ annuli (the green bars indicate the number of clean sources, the red bars above green ones show the increments by the flagged sources. See sec. 2.3 for the details).  
\label{fig03}}
\end{center}
\end{figure*}

\subsection{ Matching of the AKARI Infrared sources against the HSC optical data}

After AKARI band merging (Section 2.1), we performed source matching between the AKARI and HSC data. To identify the counterparts from each other by positional matching, a reasonable  search radius was assigned. Figure 3 summarises how we decided the radius for the source matching between the AKARI and the HSC data.

In Figure 3, the bottom left panel (Figure 3a) shows the distribution of the positional offsets of the matched sources on the  $\Delta$RA versus $\Delta$Dec plane (before the decision of the radius, we used 3$^{\prime \prime}$ as a tentative criterion). The contours in the reddish dotted lines with numbers represent the number density of the green dots normalised by the peak value at the center.
A yellow circle indicates the 3-sigma ($\sigma$) radius determined based on the  
Gaussian (magenta curves) fitted to the histograms on the top left (Figure 3b) and the bottom right (Figure 3c) panels. 
Here,  3-$\sigma$ radius corresponds to 1$^{\prime \prime}$.78, where the source density in the $\Delta$RA$cos$(dec) vs $\Delta$Dec plane approaches down to the 1 $\%$ level compared to the density peak.

Within this matching radius, we have 111,535 AKARI sources matched against the HSC optical data, which were finally divided into two groups: the clean (91,861) vs. flagged (19,674) sources 
based on the HSC flag information such as 
\texttt{base$\_$PixelFlags$\_$flag$\_$bad} (to discriminate bad pixels in the source footprint),   
\texttt{base$\_$PixelFlags$\_$flag$\_$edge} (to indicate a source is in the masked edge region or has no data),
\texttt{base$\_$PixelFlags$\_$flag$\_$saturatedCenter} (to notice that saturated pixels are in the center of a source). 
These parameters helped us when we discriminated   unreliable results with the saturated sources or ones lying at the image edge/border. 
In this work, we construct a band-merged catalogue only for the ``clean" sources, excluding the flagged ones  because the derivation of photo-z or physical modeling by SED-fitting requires accurate photometry.

The remaining 23,620 sources did not match to any HSC data (i.e., none-HSC, hereafter), some of which seem to be obscured objects in the optical bands, residing in the high-z ($>1$) universe \citep{Toba20}.

\begin{figure*}
\begin{center}
\resizebox{0.85\textwidth}{!}{\includegraphics{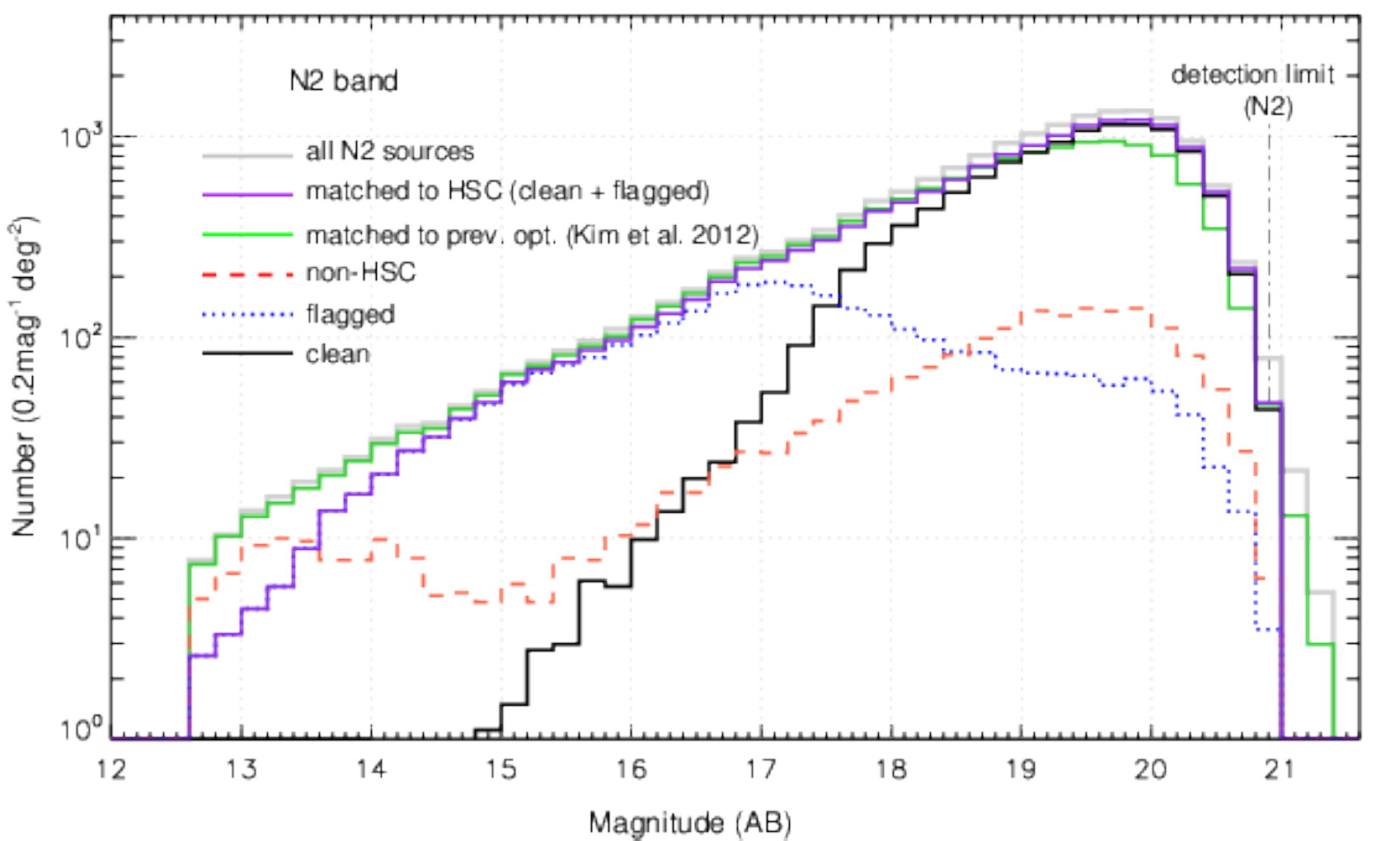}}
\caption{The number distribution of $N2$ sources as a function of magnitude (0.2 mag$^{-1}$ deg$^{-2}$). All $N2$ sources are divided into three sub-categories according to the matching results against the HSC data: clean (sold black), flagged (dotted blue), and none HSC (dashed red). The violet line shows the optically matched sources (i.e., the sum of the clean and flagged sources). The grey line shows all the $N2$ sources (i.e., the sum of the clean, flagged, and none HSC sources).
Green line represents the $N2$ sources matched to the previous optical data from the CFHT or Maidanak \citep{K12}.
\label{fig04}}
\end{center}
\end{figure*}

The histogram on the top right panel (in Figure 3d) shows the distribution of the matched sources as a function of radius interval. The green (red) bars show the number of the clean (flagged) sources  matched in each radius bin with the 0.5$^{\prime \prime}$ width. A yellow mark indicates the matching radius (1.78$^{\prime\prime}$, therefore all the sources in the 1.5--2.0 bin are matched within this radius). The  green histogram shows that half of the clean sources (48\%, 43,789) are matched within 0.5$^{\prime\prime}$ positional offsets and  34\% (31,215) are matched with the offsets between 0.5 -- 1.0$^{\prime\prime}$.   Therefore, within a 1$^{\prime\prime}$ radius, we have 82$\%$ (75,000) sources in total, matched between the AKARI and the HSC data without flagging.  Within a 1.5$^{\prime\prime}$ radius, we have 95$\%$ of the sources (87,645) matched.

\begin{figure*}
\begin{center}
\resizebox{0.95\textwidth}{!}{\includegraphics{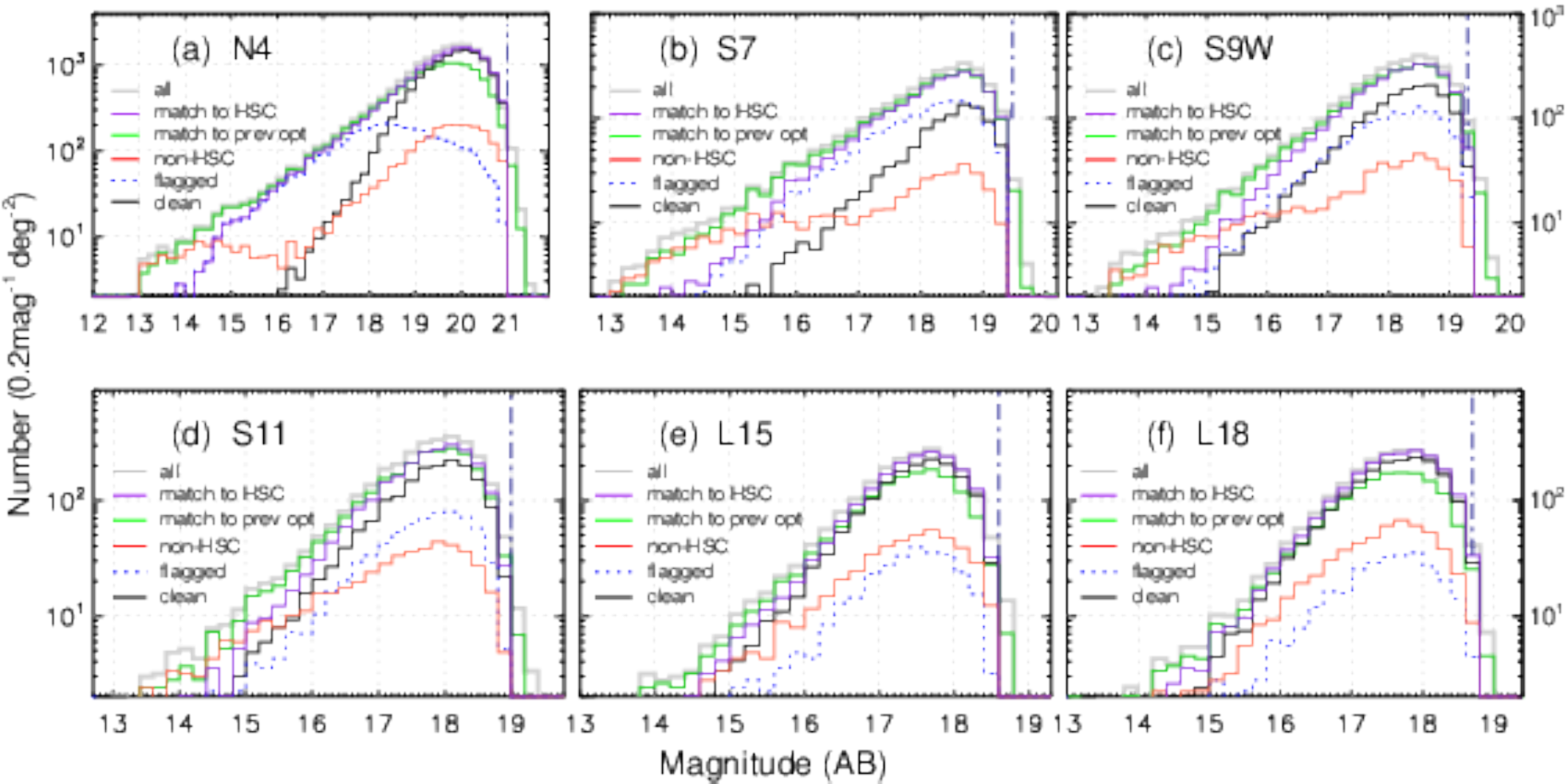}}
\caption{The number distribution of the sources as a function of magnitude, plotted in the same fashion as Figu. 4, but for the sources in the other AKARI bands. Distribution for three different sub-categories are given in the same colour as shown in Figure 4.  Vertical dot-dashed line in each panel represents the detection limit in Table 1.
\label{fig05}}
\end{center}
\end{figure*}

\begin{figure*}
\begin{center}
\resizebox{0.95\textwidth}{!}{\includegraphics{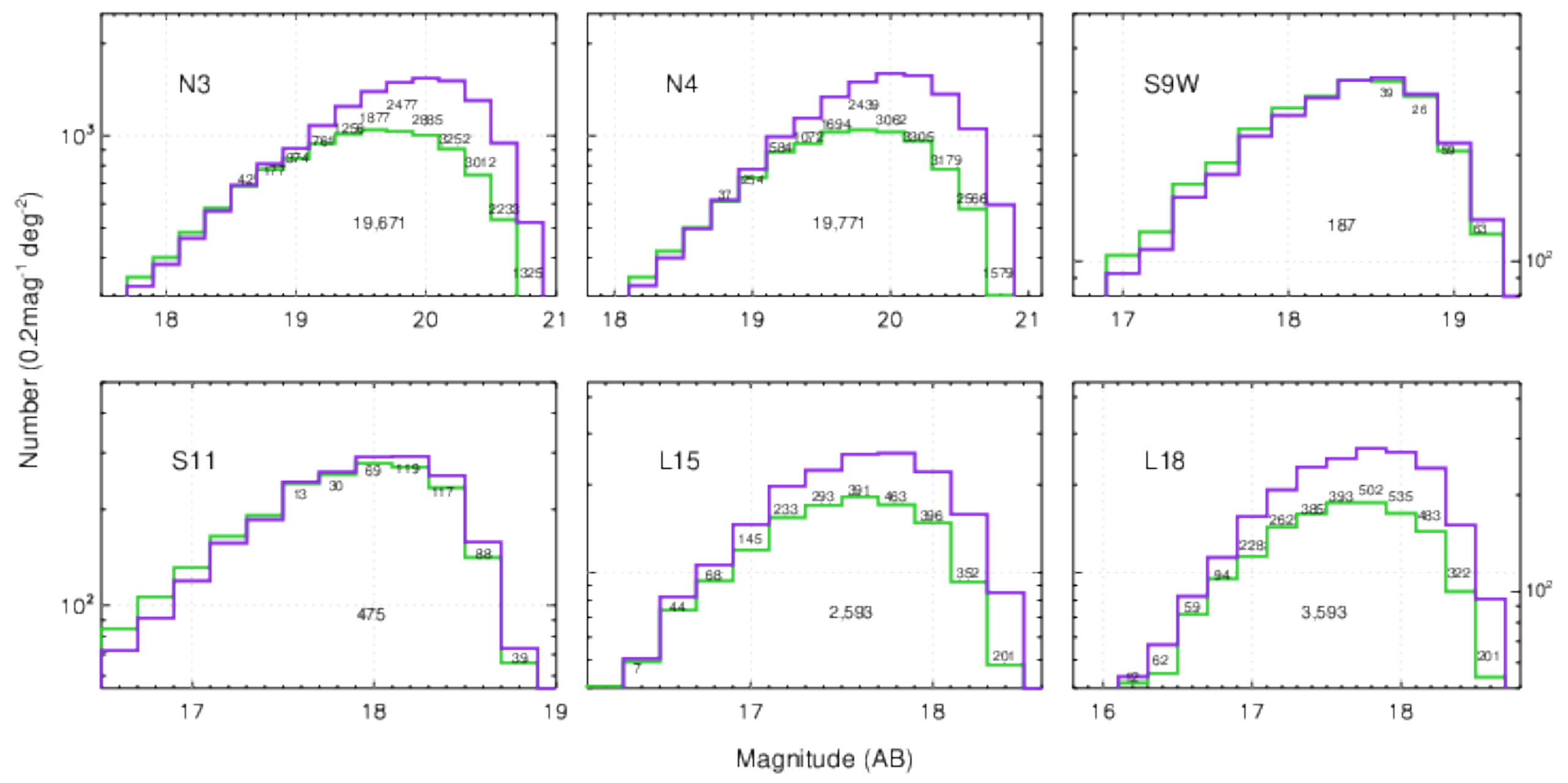}}
\caption{Selected narrow ranges from the Figure 5 near the peak of the histograms (to show the green and violet lines), which  give the comparison between the sources matched to the HSC data (violet) and to the previous optical data obtained with the CFHT and Maidanak (green).   The small numbers written on the green histogram represent the difference between the violet and green histogram in each magnitude bin, i.e., newly matched to the HSC data.
The sum of the small numbers  are presented in the middle of each panel. In this range, the violet histogram lower than green one (in the bright side)  seems mostly due to the flagged sources (blue dotted line in Figure 5) excluded in this work. 
\label{fig06}}
\end{center}
\end{figure*}

We describe the details of the matching results in Figure 4, which compares the distribution of sub-categories of the sources divided into three (clean, flagged, and none-HSC) as a function of $N2$ magnitude. A grey-coloured histogram indicates all the $N2$ sources  which corresponds to the sum of  three (clean $+$ flagged $+$ none-HSC) groups. A violet histogram shows the $N2$ sources that have HSC counterpart (clean $+$ flagged).  On the other hand, the green line shows the distribution of the $N2$ sources matched to the previous optical data (CFHT and/or Maidanak). 

In the bright magnitude range (up to 14.8 mag), there are no clean sources: all the sources are `flagged (blue dotted)' or `none HSC (red dashed)', which means that most of the bright $N2$ sources are accompanied by one of three HSC flags used to filter out the problematic sources or do not have HSC counterpart. This indicates that their HSC counterparts were affected by saturated/bad pixels, masked/edge region, or do not have valid HSC parameters, otherwise they did not match to any HSC sources. Between 13.5 and 17.5 mag, the flagged sources prevail. 
The clean sources (black line) begin to appear around 15 and go above the flagged sources at 17.5, and this predominance continues to the faint end. The N2 source counts decrease rapidly before the detection limit due to source confusion \citep{K12}. We did not include the sources fainter than the   detection limit (i.e., a small number of objects fainter than the vertical dot-dashed line   as shown in Figure 4). 
Some of the none-HSC sources have optical counterparts in the previous CFHT/Maidanak data (in the brightest range)  because they are located in the region where the $gizy$ bands (which were used for the source detection) did not cover but the CFHT/Maidanak surveys provided normal photometric measurements (see the small bump in the red dashed histogram below the green one around 13 -- 14 mag.).
These sources are beyond the concern of this work and the AKARI sources classified into flagged and/or none HSC group might be discussed later (in  separate works).

The same description for the other AKARI bands are summarised in Figure 5. The overall trends in the NIR bands is similar to $N2$: the red histogram on the top left panel (Figure 5a for the $N4$ band) shows a smaller bump around 14 mag and the other peak around 20 mag. The smaller bump is  ascribed to the unobserved area, otherwise they are probably the brightest stars (almost all the NIR sources brighter than 15 mag seem to be stars; \citep{K12}), but rejected by the pipeline or classified as flagged. 
Figure 5a also shows  there is no clean source brighter than 16 mag.  This trend in the bright end weakens/disappears as we move to the mid-IR bands. It seems that the saturation levels of the HSC bands \citep[17 - 18 mag,][]{Aihara18} correspond to the valley between two red bumps in the NIR bands, where the clean sources begin to appear.
The smaller red bump fades out in the longer wavelength (MIR) bands: it becomes weak in the $S7$ (Figure 5b), appears weaker in the $S9W$ (Figure 5c), and completely disappears in the MIR-L bands, implying the Rayleigh-Jeans tail of the stars fades out.  
In the NIR bands, the number of clean sources (black solid line) are  higher than those of flagged sources (blue dotted) near the broad peak. In the  $S9W$ bands, these two classes become comparable, while the none-HSC sources are much lower. 
In the $S9W$ and $S11$ bands, the fraction of the flagged sources (blue dotted) are between the black and the red lines. In the MIR-L bands, the blue dotted line becomes lower than none HSC sources (red line).  

While some of the bright IR sources are not fully available in this work consequentially because they are unobserved/rejected or eventually classified as flagged sources, on the other hand, there are many more fainter sources newly matched to the deeper HSC data as shown in Figure 6 (for example,  19,771 at $N4$, and 2,593 at $L15$, respectively).  Only small ranges were presented in the figure: note the range where the violet histogram is higher than the green one. The number in each magnitude bin indicate the sources newly matched to the HSC data. 

However, the bigger red bumps (in Figure 5) near the faint ends  indicate our HSC survey  was still not   deep enough to identify all the faint IR sources,  which left some of the AKARI sources unmatched against the HSC data.  
On the AKARI colour-colour diagrams (in sec. 4), these faint IR sources are located in the same area as the other (optically identified) sources, which implies that they have the similar IR properties. They are probably infrared luminous SFGs, but appear to have dropped out in the HSC bands. Seemingly,   they might be a certain kind of highly obscured dusty systems in high-z. A detailed discussion based on the selected sample is presented in \cite{Toba20}.

\begin{table*}
	\centering
	\caption{Summary of the multiwavelength data sets: the detection limits }
	\label{tab1}
	\begin{tabular}{ccccc}
	\hline
	\hline
	Data   & Band &  Effective wavelength & (5$\sigma$) detection limit   \\
                &     &  ($\mu$m) & AB / $\mu$Jy    \\

	\hline
	                  & $N2$   &   2.3  &  20.9 / 15.4    \\
                      & $N3$   &   3.2  &  21.1 / 13.3    \\
    AKARI/IRC         & $N4$   &   4.1  &  21.1 / 13.6    \\
 NEP-Wide Survey      & $S7$   &   7    &  19.5 / 58.6    \\
5.4 deg$^2$           & $S9W$  &   9    &  19.3 / 67.3    \\
\citep{K12}     & $S11$  &   11   &  19.0 / 93.8    \\
	                  & $L15$  &   15   &  18.6 / 133   \\
	                  & $L18W$ &   18   &  18.7 / 120   \\
	                  & $L24$  &   24   &  17.8 / 274   \\
 \hline
      	              &  $g$   &   0.47 &  28.6 / 0.01   \\
    Subaru/HSC        &  $r$   &   0.61 &  27.3 / 0.04   \\
 5.4 deg$^2$          &  $i$   &   0.76 &  26.7 / 0.08   \\
  \citep{Oi20}        &  $z$   &   0.89 &  26.0 / 0.14   \\
		              &  $y$   &   0.99 &  25.6 / 0.21   \\
  

\hline
CFHT/MegaPrime &\multirow{2}{*}{$u$} &\multirow{2}{*}{0.36} & \multirow{2}{*} {25.4 / 0.25  }   \\
3.6deg$^2$\citep{Huang20} &   &   &    \\

 \hline
                      & $u^{*}$ &   0.39 &  26.0 / 0.16   \\
CFHT/MegaCam$^{\rm a}$    &  $g$    &0.48 &  26.1 / 0.13   \\
2 deg$^2$ \citep{H07}&$r$&  0.62 &  25.6 / 0.21   \\
0.7 deg$^2$ \citep{Oi14} &$i$&  0.75 &  24.8 / 0.39   \\
	 	              &   $z$   &   0.88 &  24.0 / 0.91   \\
 \hline	 	          
 Maidanak/SNUCam      &   $B$  &   0.44 &  23.4 / 1.58   \\
4 deg$^2$ \citep{J10} &$R$&  0.61 &  23.1 / 2.09   \\
         	          &   $I$  &   0.85 &  22.3 / 4.36   \\
 \hline
  KPNO/FLAMINGOS        &   $J$   &  1.2  &  21.6 / 8.32   \\
5.1 deg$^2$ \citep{Jeon14} &$H$&  1.6  &  21.3 / 10.96  \\

 \hline 
CFHT/WIRCam             &   $Y$   &  1.02 &  23.4 / 1.58   \\
0.7 deg$^2$ \citep{Oi14} & $J$  &  1.25 &  23.0 / 2.29   \\
    	                &   $K_S$ &  2.14 &  22.7 / 3.02   \\
 \hline
 Spitzer/IRAC                   & IRAC1 & 3.6 & 21.8 / 6.45  \\
7 deg$^2$ \citep{Nayyeri18}  & IRAC2 & 4.5 & 22.4 / 3.95  \\
\multirow{2}{*}{0.4 deg$^2$ \citep{Jarrett11}}& IRAC3 & 5.8 & 20.3 / 27.0  \\
                                & IRAC4 &  8  & 19.8 / 45.0  \\
\hline	
                       &    W1     & 3.4  & 18.1 / 18     \\
    WISE               &    W2     & 4.6  & 17.2 / 23     \\
\citep{Jarrett11} & W3& 12  & 18.4 / 139    \\
   	                   &    W4     & 22   & 16.1 / 800    \\

\hline
 Herschel/PACS$^{\rm b}$ & Green &  100 & 14.7 / 4.6 mJy     \\
0.44 deg$^2$ \citep{Pearson19} & Red   & 160 & 14.1  / 8.7 mJy     \\

\hline
Herschel/SPIRE$^{\rm c}$  &  PSW  &  250  &  14 / 9.0 mJy   \\
9 deg$^2$ \citep{Pearson17}&  PMW  &  350  &  14.2 / 7.5 mJy    \\
                                  &  PLW  &  500  &  13.8 / 10.8 mJy   \\
\hline
SCUBA-2/NEPSC2$^{\rm d} $  & \multirow{2}{*}{850} & \multirow{2}{*}{850} & \multirow{2}{*} {1.0 - 2.3 mJy  }   \\
2 deg$^2$ \citep{Shim20} &   &   &    \\

\hline
\end{tabular}

(a) The detection limits refer to the 4$\sigma$ flux over a circular area with a diameter of 1$^{\prime\prime}$. 
(b) The detection limits refer to 3$\sigma$ instrumental noise sensitivities.
(c) The detection limits refer to the Open Time 2 (OT2) sensitivity. 
(d) The detection limits refer to the 1-$\sigma$ rms noise  (or 4.7-11 mJy at 80\% completeness).

\end{table*}

\section {Complementary Data Sets}

After the identification of $AKARI$ IR sources with the HSC optical data, we used all available photometric catalogue/data over the NEPW field to construct multi-band catalogue. In this section, we briefly describe the data sets used in this work.  Just as  Figure 1 showed the coverages of various surveys,  Table 1 and Figure 7 summarise the photometric bands and depths of the surveys.  Figure 7a also shows why source detection changes in different instrument/filter systems.

\subsection{Ancillary Optical Data: CFHT and Maidanak }
 
It is not easy to take the entire 5.4 deg$^2$ area in a uniform manner unless we have a large-FoV instrument with an appropriate filter system covering a good enough wavelength range. This was what made our previous optical surveys divided into two different data sets 10 years ago: one obtained with the MegaCam ($u^{*}$, $g$, $r$, $i$, $z$) on the central 2 deg$^2$ area, and the other with SNUCAM $B$, $R$, and $I$  \citep{Im10} of the Maidanak observatory over the remaining part of the NEPW field. The detailed description of these two surveys can be found from \cite{H07} and \cite{J10}, respectively. However, the western half  of the central CFHT field was not observed by $u^*$ band. Also, due to the different filter systems and  depths between these two optical surveys, homogeneous analysis with optical counterparts over the whole field was practically impossible. 
Another optical survey was carried out later \citep{Oi14} on the NEPD field ($\sim$0.7 deg$^2$) and finally provided MegaCam $u^*$ data for the western half area as well as the supplementary WIRcam data ($Y, J,  K_{s}$). 
In addition, the CFHT MegaPrime $u$-band observation was performed over a 3.6 deg$^2$ area on the eastern side of the NEPW field \citep{Huang20}\footnote{http://doi.org/10.5281/zenodo.3980635} to replenish the insufficient (central 2 deg$^2$ only; see Figure 1) coverage of the MegaCam $u^*$-band.  Because how to calibrate photo-z is a significant issue under the circumstances that a huge number of sources remain without redshift information,  availability of u-band data is crucial to check the UV extinction properties and to improve photo-z accuracy \citep{Ho20}. 
We combined all these supplementary optical data: a small systematic   shift of WCS ($< 1^{\prime\prime}$) in each optical data with respect to the HSC were corrected first, and the matching radii for each data were decided based on the mean positional differences (see Figure 8).  
The number of sources matched to the HSC data are summarised in Table 2.

\begin{figure*} 
\begin{center}
\resizebox{0.95\textwidth}{!}{\includegraphics{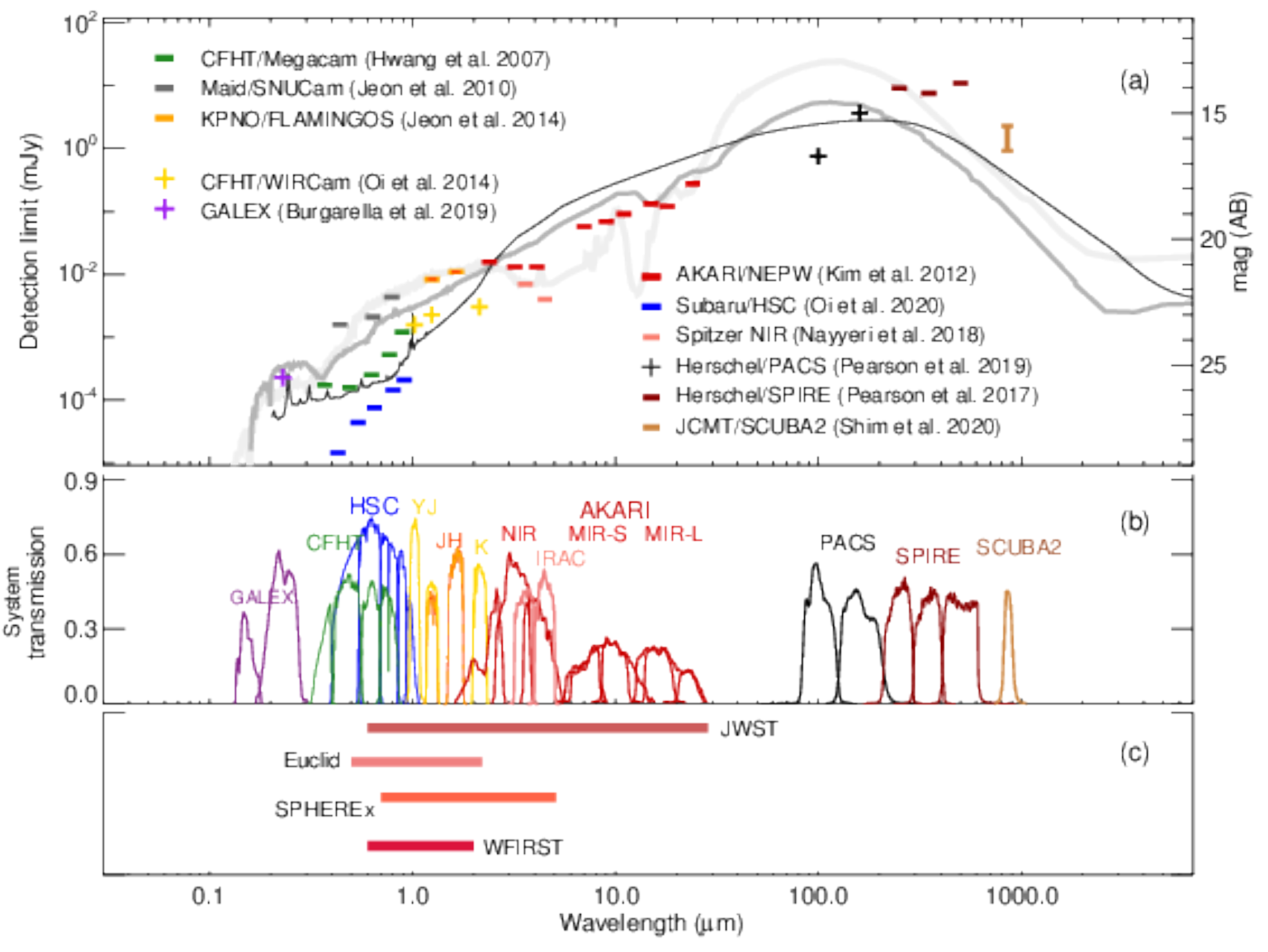}}
\caption{(Top) The depths of various surveys with different instruments/filter systems over the NEP, in terms of the 5-sigma limiting magnitude. The cross symbols (\textbf{+}) imply that the survey was dedicated only to the NEPD field. Typical templates \citep{Polletta07} are given to show how a local ULIRG (e.g., Arp220-type at z=0.3, faintest thick line), a type-1 Seyfert (at z=0.6, dark grey), or a dusty torus model (at z=1.0, black thin line) looks in this plot. All of them are normalised at $N2$ detection limit. (Middle) The system transmission/filter shapes are presented.  (Bottom) The comparison of the spectral range in the IR  to be covered by the future space missions is presented, as shown by the horizontal bars. 
\label{fig07}}
\end{center}
\end{figure*}


\begin{figure*}[h]
\includegraphics[width=0.33\textwidth]{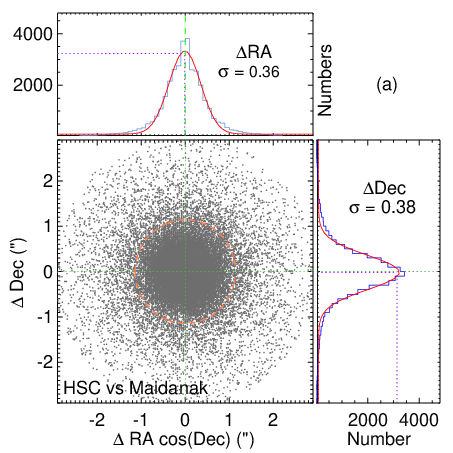}    
\includegraphics[width=0.33\textwidth]{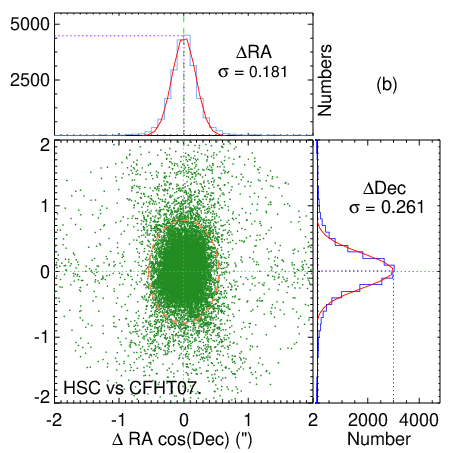}    
\includegraphics[width=0.33\textwidth]{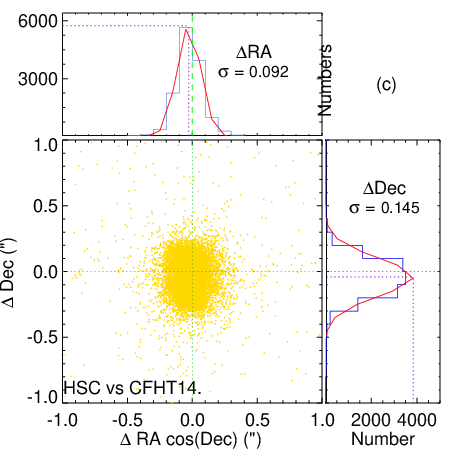}    
\includegraphics[width=0.33\textwidth]{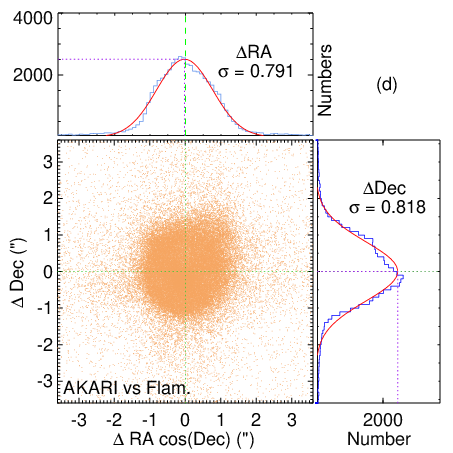}    
\includegraphics[width=0.33\textwidth]{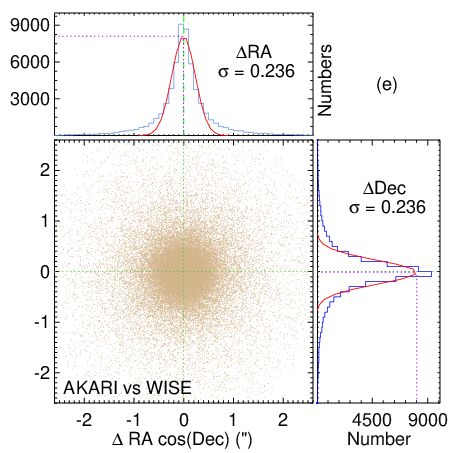}    
\includegraphics[width=0.33\textwidth]{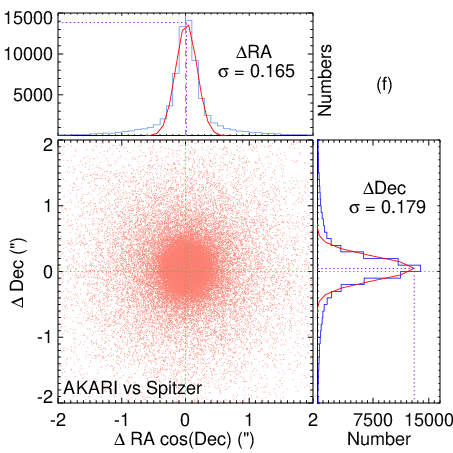}    
\includegraphics[width=0.33\textwidth]{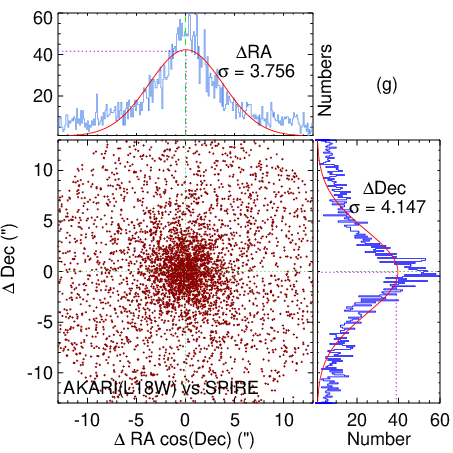}   
\caption{Some examples of the astrometry-offset distributions of the sources matched between the clean (HSC-AKARI)   and  the other supplementary data.  The matching circles (or ellipses) were decided based on the representative value of the offset (width of the offset histogram), 3-$\sigma$ derived from the Gaussian fit to the histogram of $\Delta$RA and $\Delta$Dec. Most of them appear to be circular shape except for the matching against the CFHT data. (a) HSC vs Maidanak (b) HSC vs CFHT (2007) (c) HSC vs CFHT (2014) (d) AKARI vs FLAMINGOS (e) AKARI vs WISE (f) AKARI vs Spitzer (g) AKARI (MIR-L) vs SPIRE.
\label{fig08}}
\end{figure*}

\begin{table*}
	\centering
	\caption{Summary of the Matching against the Supplementary Data }
	\label{table2}
	\begin{tabular}{cccccc}
	\hline
	\hline
\multirow{2}{*}{Main Data}&Supplementary Data &3-$\sigma$ Radius & PSF Size & Number of&\\
         & (Reference) & ($^{\prime\prime}$) & (FWHM, $^{\prime\prime}$) & matched sources &\\
	\hline
\multirow{5}{*}{HSC}& Maidanak/SNUCam \citep{J10} &  1.14    & 1.1 - 1.4 & 33,485 & \\
{Subaru}   & CFHT/MegaCam \citep{H07}  &0.54/0.78$^{\rm a}$ & 0.7 - 1.1 & 23,432 & \\
        & CFHT/Mega-WIR \citep{Oi14} &0.28/0.44$^{\rm a}$ & 0.8 - 0.9 & 15,261 & \\
     & CFHT/MegaPrime-u \citep{Huang20} &0.43/0.55$^{\rm a}$ & 0.8 - 1   & 31,851 & \\
      & GALEX \citep{Burgarella19}        &  3.2    &    5.0    & 58 & \\ 
  \hline
\multirow{6}{*}{AKARI}  & KPNO/FLAMINGOS \citep{Jeon14} &  2.2  & 1.7 - 1.8 &  46,544 \\
      & WISE \citep{Jarrett11}      &  0.9 &  $\sim$ 6  &  60,062 \\
           & Spitzer \citep{Nayyeri18}    &  1.2  & 1.78      &  79,070 \\
    &PACS \citep{Pearson19} & 3.6/ 6.3$^{\rm b}$ & 6.8/ 11.3 & 882/ 463 \\
    &SPIRE \citep{Pearson17}  &  8.1$^{\rm c}$   &  17.6    &  3,109 \\
\hline
\hline
\end{tabular}\\
(a) The matching radii along the RA and Dec are not the same (see Figure 8). 
(b) The radii for the 100 $\mu$m and 160 $\mu$m band, respectively.   
(c) The source extraction was done on the 250 $\mu$m map, and the sources were catalogued with photometry in all three SPIRE bands. 
\end{table*}

\begin{figure*}
\begin{center}
\includegraphics[width=0.95\textwidth]{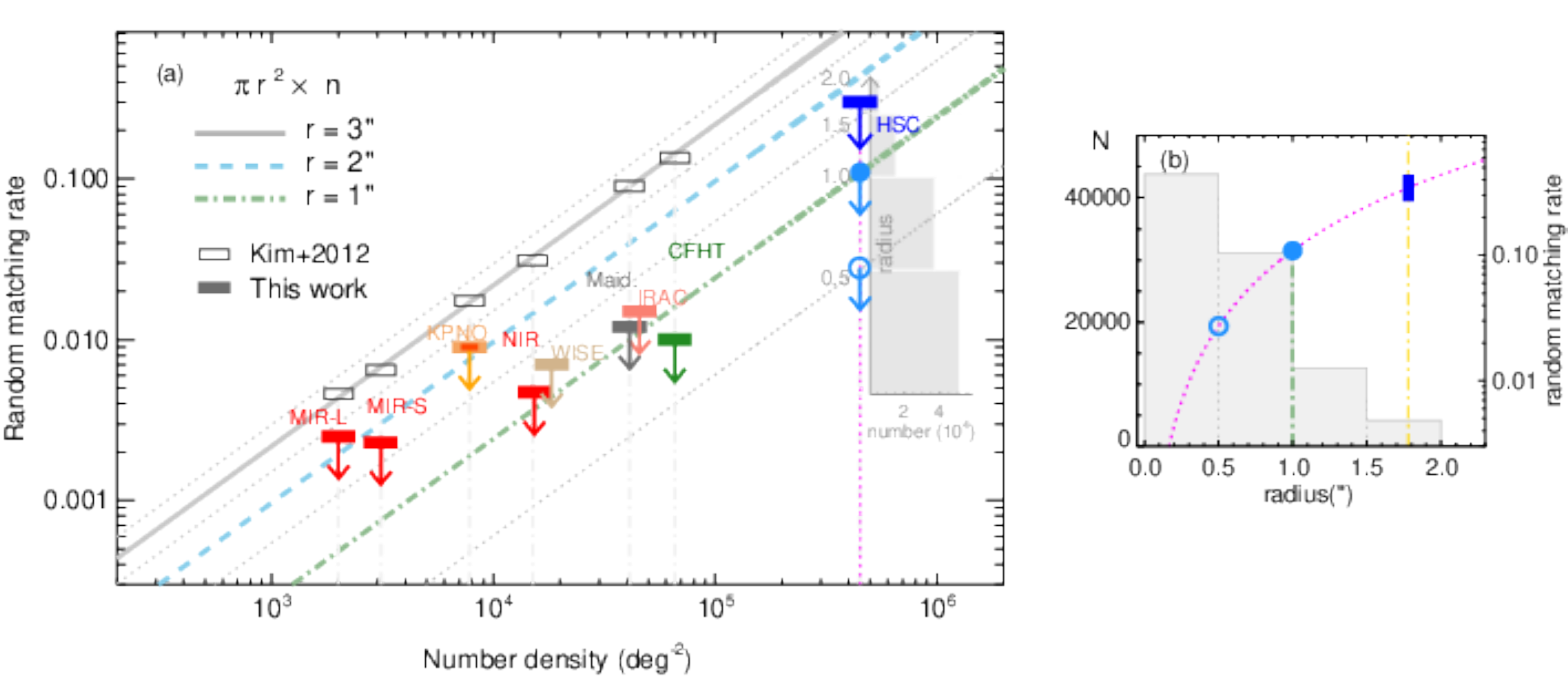}
\caption{(Left) Random matching rate in terms of the source density and matching radius. The open boxes are taken from the Fig 14 in \citet{K12}, which represent the actual tests, i.e., number counts of the sources randomly matched to each data using 3 $^{\prime\prime}$ radius. The grey thick line shows these measurements are described by a simple relation ($n\pi r^{2}$). If we use 2$^{\prime\prime}$ or 1$^{\prime\prime}$ radius, the random matching rate will be decreased as described by cyan dashed and green dot-dashed lines. The faint dotted lines between them  indicate 0.5$^{\prime \prime}$ increments. The filled boxes in different colours show the random matching estimates  when a certain source is matched against the data with our radii determined in this work. The grey histogram is taken from Fig 3d. (Right) The embedded grey histogram on the left panel can be plotted with random matching rate as shown in the right panel. The magenta curve shows the random matching rate when the number density of source is 4.5$\times 10^{5}$ (which corresponds to the magenta vertical line in the left panel).  Vertical yellow line represents the matching radius from Figure 3. }
\label{fig09}
\end{center}
\end{figure*}

\subsection{Spectroscopic and Photometric Redshifts } 

Following the 
optical identification of the AKARI sources with the deep HSC data, we incorporated all available spectroscopic redshifts (spec-z) data  to the clean  AKARI-HSC sources (therefore, the redshift information matched to the flagged sources are not included here.). 
There have been many spectroscopic observations over the AKARI's NEP area. The most extensive and massive campaign covering the entire NEPW field was done by \citet{Shim13}: they targeted the NEPW sources selected primarily based on the MIR fluxes at 11 $\mu$m ($S11<18.5$ mag) 
and at 15 $\mu$m ($L15<17.9$ mag) 
to see the properties of the MIR selected SFGs. 
Most of these flux-limited sources turned out to be various types of IR luminous populations of galaxies.  A smaller number of secondary targets (35$\%$ out of their targets) are also selected to catch some rare types of objects, such as obscured AGNs, BzKs, super cluster candidates, etc. They provided the spectra of 1796 sources (primary targets: 1155, secondary targets: 641), and the redshifts for 1645 sources were measured.  These spectroscopic sources are classified into several types (e.g., star, type1 AGN, type2 AGNs, galaxy, unknown).  Recently, a new spectroscopic campaign over the whole NEPW area with the MMT/Hectospec has been initiated to carry out a homogeneous survey for the 9 $\mu$m selected galaxies (MMT2020A/B, PI: H. S. Hwang). 

We also took the redshift/type information from many other spectroscopic surveys on the NEPD field.  For example, Keck/DEIMOS observations were conducted in order to measure the spectroscopic redshift and calibrate photo-zs for MIR galaxies (DEIMOS08) \citep{Takagi10}, and to measure [OII] luminosity against 8$\mu$m luminosity (DEIMOS11) \citep{Shogaki18},   and more recently, to check the line emission evidence of AGNs and metallicity  diagnostics of SFGs, etc.  (DEIMOS14, DEIMOS15, DEIMOS17) \citep{HKim18}.  
Another series of spectroscopic observations with Gran Telescope Canarias (GTC)/OSIRIS were carried out  between 2014 and 2017 (e.g., GTC7-14AMEX, GTC4-15AMEX, GTC4-15BMEX, and GTC4-17MEX) \citep{DiazTello17} to see the X-ray signatures of highly obscured and/or Compton-thick (CT) AGNs   along with the identification by the \textit{Chandra} data \citep{Krumpe15}.
Subaru/FMOS spectroscopy was also obtained  to investigate the mass-metallicity relation of IR SFGs in the NEP field \citep{Oi17}. \citet{Ohyama18}   provided the polycyclic aromatic hydrocarbons (PAHs) galaxy sample with redshift measurements through the SPICY projects done by AKARI/slitless spectroscopy.  We combined all these redshift information.

Using these spectroscopic redshifts as a calibration sample, \citet{Ho20} estimated photo-zs using the photometry from $u^{*}$-band to the NIR band (IRAC2 and/or WISE2). 
They checked photometry by comparing colours/magnitudes to discriminate the unreasonable data so that they could obtain reliable results when they use the software \texttt{Le PHARE}. After the photo-zs were assigned,  they presented effective star-galaxy separation scheme based on the $\chi^2$ values.

\subsection{Supplementary Near-/Mid-IR Data}

While the optical data are crucial for identifying the nature of the corresponding AKARI sources, 
the $J$-, $H$-, and $K$-band data are useful to bridge the gap in wavelength  between optical $y$ and $N2$ band. The CFHT/WIRCam  covered a limited area, but provided useful $J$- and $K$-band photometry \citep{Oi14}.  The NIR ($J$, $H$) survey that covered almost the entire NEPW area ($\sim$ 5.2 deg$^{2}$) was done by  FLAMINGOS of the Kitt Peak National Observatory (KPNO) 2.1 m telescope although the depth is shallower than WIRcam data
(see Figure 1 and 7).
For  complementary photometry in the near- to mid-IR, we also included the publicly available data taken by  Spitzer \citep{Werner04} and WISE\footnote{Also see https://wise2.ipac.caltech.edu/docs/release/allwise/expsup/sec2\_1.html} \citep{Wright10}.
The  catalogue by \cite{Nayyeri18} provides 380,858 sources covering the entire NEPW field ($\sim$ 7 deg$^2$)  with higher sensitivity (21.9 and 22.4 mag at the IRAC1 and IRAC2, respectively) and slightly better spatial resolutions compared to the $N3$ and $N4$, which are useful to cross check against the longer wavelength data having larger PSFs (e.g., the SPIRE or SCUBA-2 data).

\subsection{FIR/Smm Data from the Herschel and SCUBA-2} 

Herschel carried out the 0.44 deg$^2$ and 9 deg$^2$ surveys over the NEP field with the Photoconductor Array Camera and Spectrometer \citep[PACS:][]{Poglitsch10} and Spectral and Photometric Imaging REceiver instrument \citep[SPIRE:][]{Griffin10}, respectively.
From the PACS NEP survey \citep{Pearson19, Burgarella19}, the Green (100 $\mu$m) and Red (160 $\mu$m) bands provide 1380 and 630 sources over the NEPD field, with the flux densities of 6 mJy and 19 mJy at the 50\% completeness level, respectively.
The SPIRE also carried out NEP survey as an open time 2 program (PI: S. Serjeant, in 2012), and completely covered the entire NEPW field, at the 250, 350, and 500 $\mu$m (achieving 9.0, 7.5, and 10.8 mJy sensitivities at each band). Source extraction was carried out on the 250 $\mu$m map, and approximately $\sim$4800  sources were catalogued with the photometry in all three SPIRE bands. The more detailed description of the data reduction and photometry can be found in \cite{Pearson17}.
Compared to the optical or NIR data, the Herschel (PACS or SPIRE) data have larger positional uncertainties with much larger PSF sizes. This can make the identification of sources against the AKARI  data potentially ambiguous when we carry out the positional matching, even though the radius was determined reasonably (the 3-sigma radii are smaller than the PSF sizes, in general, as shown in Table 2).
In our catalogue, the cases that multiple AKARI clean sources are lying within the searching radius around the SPIRE/PACS positions  were not included so that we clearly chose only one AKARI counterpart against the Herschel sources. 
The 850 $\mu$m submilimetre (sub-mm) mapping on the NEPW field is currently ongoing  by one of the large programs with the JCMT/SCUBA-2 \citep{Shim20}. They released a mosaic map and a catalogue for the central 2 deg$^2$ area first.  They provide 549 sources above 4-$\sigma$ with a depth of 1.0--2.3mJy per beam.  The source matching against the AKARI-HSC clean catalogue was carried out based on the likelihood ratio.  We derived the probability of counterpart for a 850$\mu$m source, using both the magnitudes distribution of IRAC1 and IRAC2 bands (which are deeper than those of the AKARI NIR bands) and their colour as well as those of three SPIRE bands.  We took 46 sources as robust AKARI counterparts for the 850 $\mu$m sources because they are matched to both IRAC and SPIRE with high (95$\%$) probability. We also included 16 sources as decent AKARI counterpart   because in these examples there is only one IRAC/SPIRE source within the 850$\mu$m beam. Lastly, 4 sources matched to IRAC with high probablity but we are uncertain about the SPIRE cross-identification. 
However, when multiple optical sources were associated with any given SPIRE or SCUBA-2 source,  if  real optical counterpart was already classified as flagged sources,  then it could be  complicated/a potential issue.

\section{The properties of the AKARI sources Identified by HSC Survey Data }

\begin{figure*}
\begin{center}
\resizebox{0.9\textwidth}{!}{\includegraphics{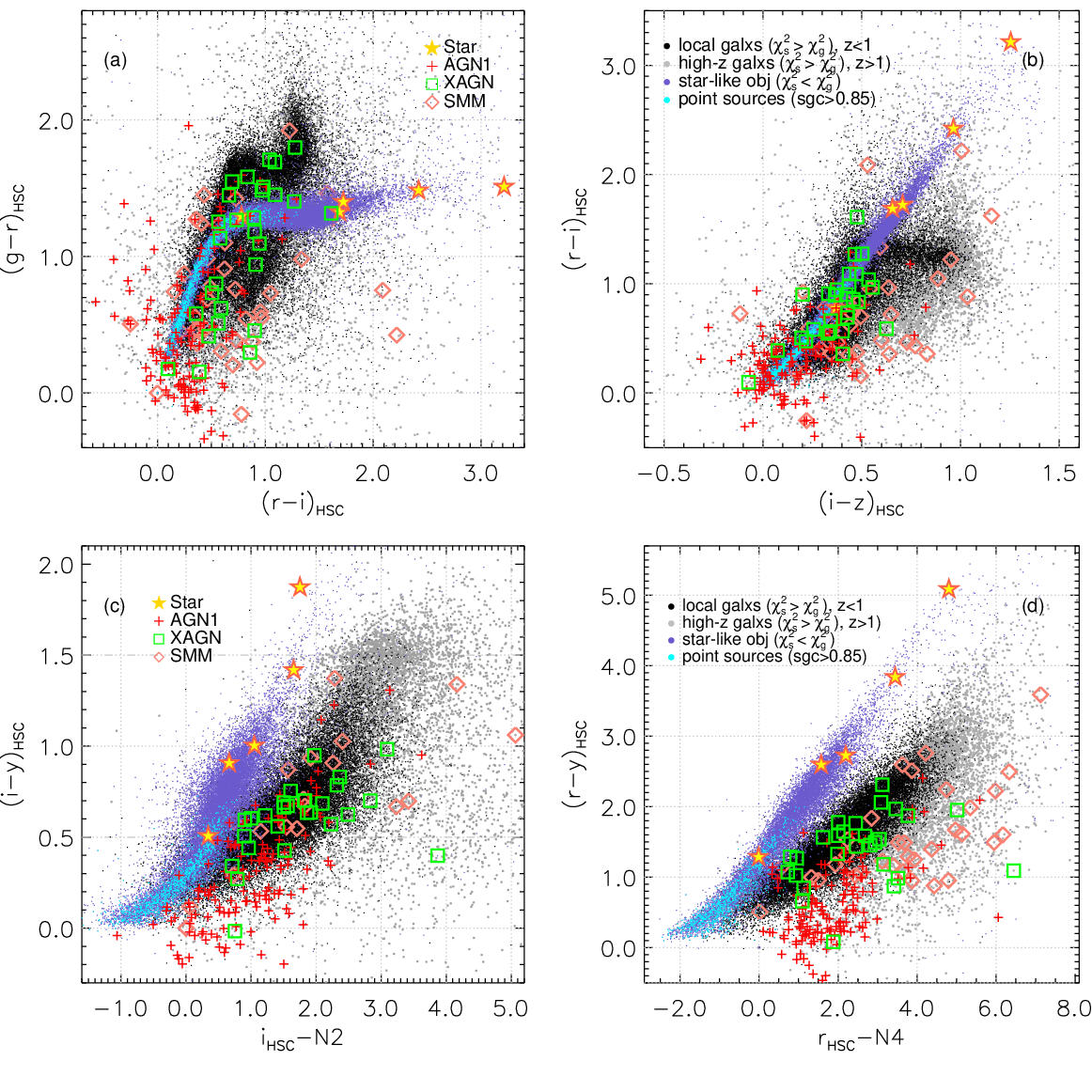}}
\caption{Colour-colour diagrams based on the HSC optical and AKARI NIR bands. Violet dots represent the sources classified as star-like sources, and black dots represent the extra-galactic sources with z$_{\rm phot} < 1$ while the grey dots are the source  with z$_{\rm phot} > 1$  \citep{Ho20}. Cyan dots are high-stellarity (sgc$>0.8$) sources. Yellow stars represent the Galactic stars observed by  the spectroscopic survey \citep{Shim13}.  Red crosses are  AGNs (type1), also confirmed by \citet{Shim13}. Green boxes are  AGNs that have X-ray data \citep{Krumpe15,DiazTello17}. Salmon diamons are galaxies observed by the SCUBA-2 survey \citep{Shim20}. All axes are in units of AB mag. }
\label{fig10}
\end{center}
\end{figure*}

\begin{figure*}
\begin{center}
\resizebox{0.9\textwidth}{!}{\includegraphics{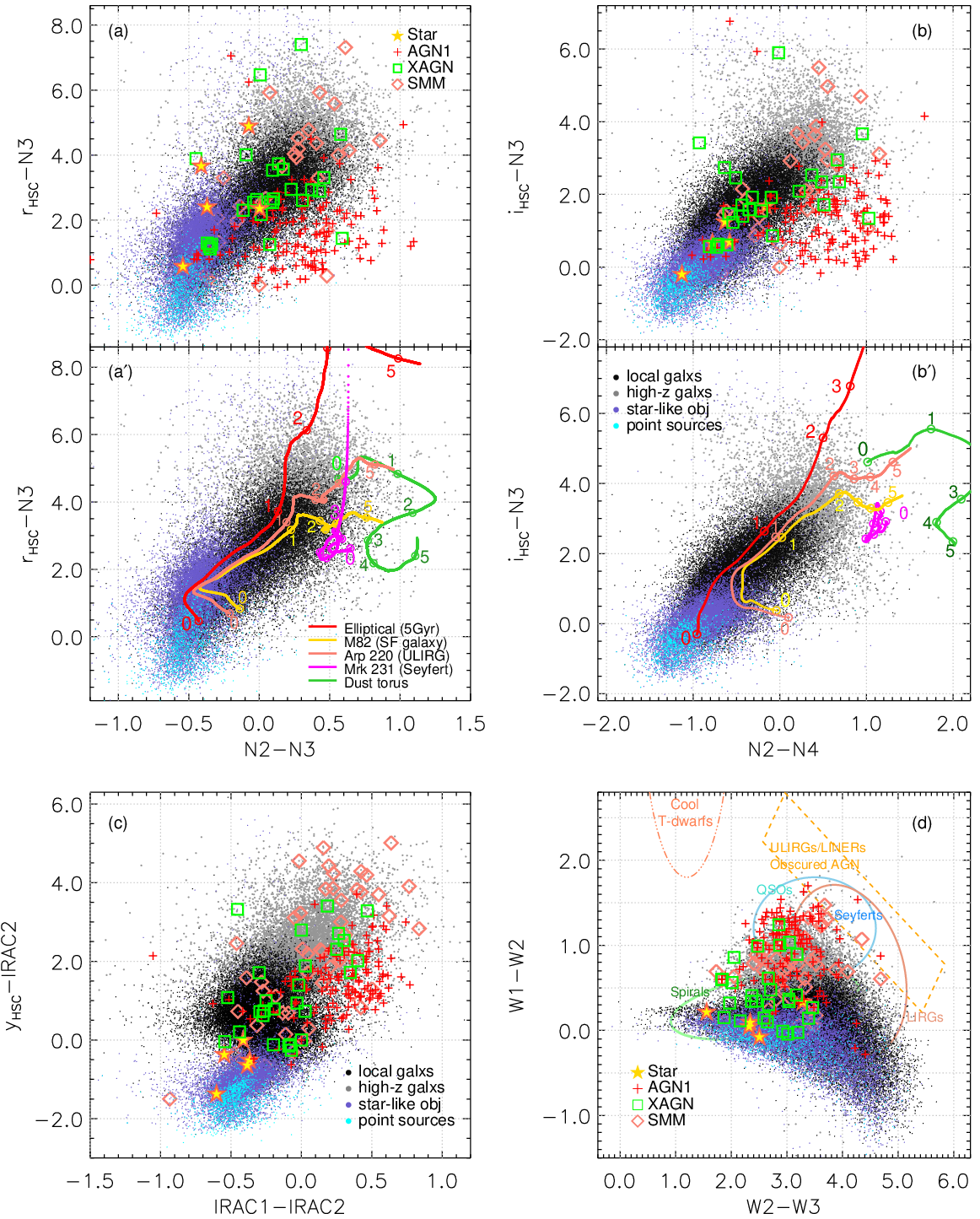}}
\caption{Colour-colour diagrams based on the HSC and various NIR bands (AKARI, Spitzer, and WISE), with the symbols in the same fashion as presented in Figure 10: star-galaxy separation and how their locations change in different colour-colour diagrams are described. For top panels (a and b), we present the evolutionary tracks of several model templates from \citet{Polletta07}.  All axes are in units of AB mag except for the lower right panel (d), which is given in units of Vega mag  to compare with the diagram from \citet{Wright10}. The mag offsets ($\Delta m$) between AB and Vega system ($m_{\textrm{AB}} = m_{\textrm{Vega}} + \Delta m$) are 2.70, 3.34, 5.17 mag for W1, W2, and W3, respectively. } 
\label{fig11}
\end{center}
\end{figure*}

\begin{figure*}
\begin{center}
\resizebox{0.9\textwidth}{!}{\includegraphics{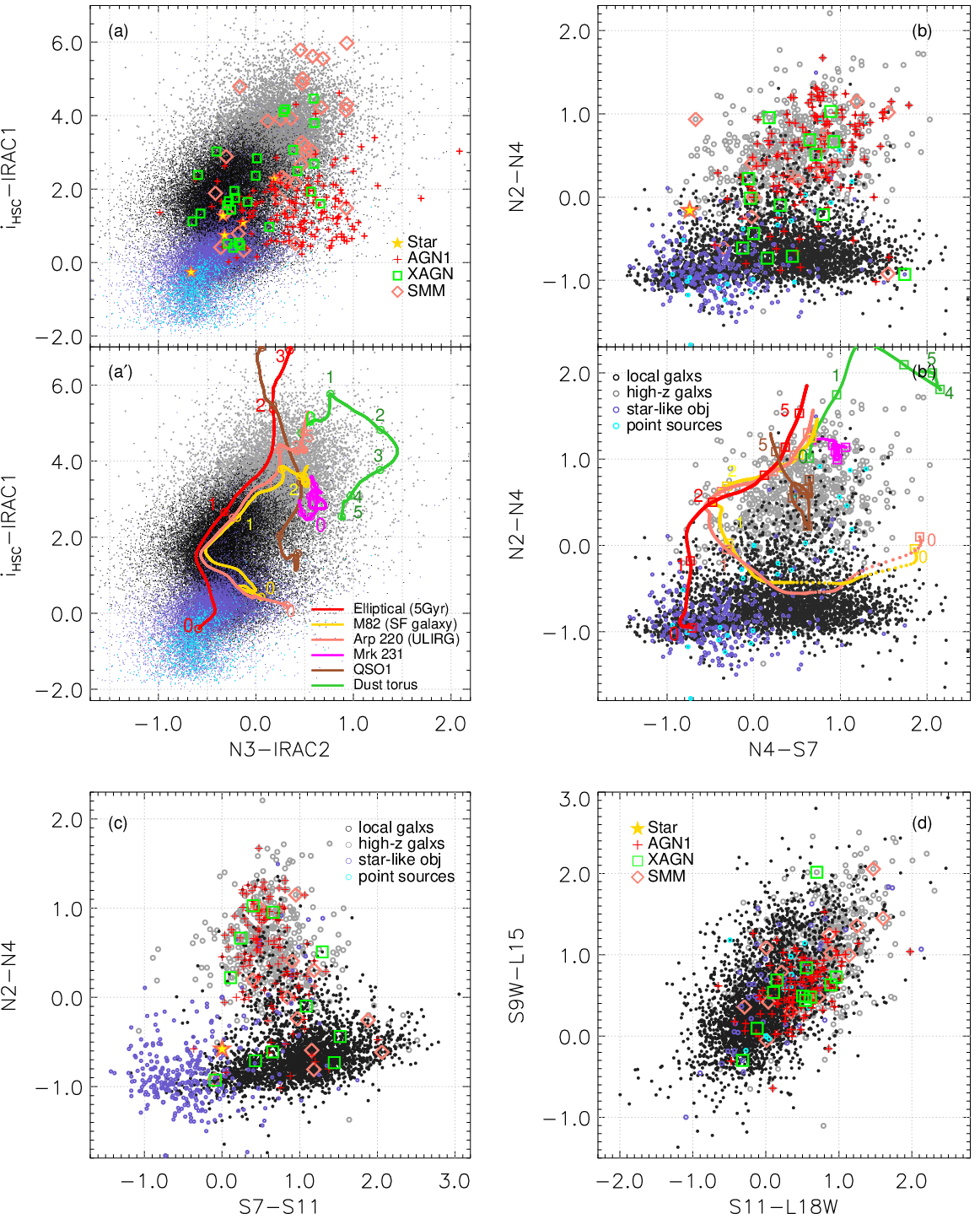}}
\caption{Colour-colour diagrams based on the various combination of NIR to MIR bands, in the same fashion as presented in Figure 10 and 11. Stars begin to fade out in MIR bands. For top panels (a and b), we also present the evolutionary tracks (a$^{\prime}$ and b$^{\prime}$ of several model templates from \citet{Polletta07}. All axes are in units of AB mag }
\label{fig12}
\end{center}
\end{figure*}

\subsection{Overview: reliability vs random matching } 
When we seek to find a counterpart against a certain data using a searching radius ($r$), there is always a possibility that we could encounter a random source (which is not associated with real counterpart) inside the circle defined by the radius ($\pi r^{2}$). 
This is the probability that a source can be captured simply by this circular area at any random position of the data.
The higher the source density of data we match against, the higher the random matching rate becomes. If we use a larger radius, then the probability is also increased.
\cite{K12} showed that it can be expressed in terms of number density ($n$) of data and cross-section. See the open boxes and grey solid lines in Figure 9a, which are taken from the Figure 14 in \cite{K12}.

Practically, these have to be regarded as upper limits of a false match possibilities (the worst case),  because, in general, two data sets for source matching are  correlated with each other: these data are obtained from the same field of view,  not from two different arbitrary sky positions. Therefore, the matching results are generally better than this estimation (as indicated by the downward arrows). 

To give how to interpret the source matching in a consistent way with \cite{K12},   we used the same analyses,  using our matching radii determined in sections 2 and 3, and compared with the plots in the previous work.
In \cite{K12}, they used the uniform radius (3$^{\prime\prime}$, the open boxes in Figure 9a). In this work, however, the matching radii were chosen  based on the mean positional offset of the matched sources between the data. This gives much smaller radii compared to those from \cite{K12}, even smaller than the PSF sizes which are also used frequently for matching criteria. 
In Figure 9a, the highest number density of the HSC data ($\sim 4.5\times10^{5}$ sources per deg$^{2}$) seems to have a high random matching (the filled blue box). Here, we should remind that only 5\% of the sources are matched outside 1.5$^{\prime\prime}$ and  82\% of the sources are matched within 1$^{\prime\prime}$, as shown by the grey background histogram, which is the same as Figure 3d (but x- and y-axis are transposed/rotated here). Therefore, the random matching rate is  better than the filled circle on the green dot-dashed line. 
On the right panel (b) in Figure 9, this grey histogram is re-plotted with random matching rate, which gives more straightforward description.
However, it should be noted that this is just an estimation of the probability when the test was performed on random positions.  Only a small fraction will suffer from the random matching, in reality.
The sources matched within  0.$^{\prime\prime}$5 seem obviously reliable (downward below the open circle). In the same fashion,  the matching with the other supplementary data (in green, grey, salmon boxes, and so on), which have a much lower number density, are less affected by random matching and relatively safe compared to the HSC data.

\subsection{Colour-colour Diagrams} 
 
We describe the photometric properties of the NEPW infrared sources matched to the HSC optical data using various colour-colour diagrams (Figures 10 to 12). The colour-colour diagrams are also helpful to see, from a statistical standpoint, if the source matching was accomplished properly. In each diagram, we used several different colours and symbols to distinguish between the  different types of sources. Violet dots indicate the sources classified as star-like, which were fitted better to the stellar model templates rather than  the  galaxy templates,  following the diagnostics ($\chi^{2}_{\rm star} <\chi^{2}_{\rm galaxy}$) by \cite{Ilbert09}    when the photo-z estimation was performed with \texttt{Le PHARE} \citep{Ho20}.
The sources fitted better to the galaxy templates are divided into two groups by redshift: black dots represent the local (z$_{\rm phot}<1$) galaxies and the grey dots represent the high-z ($>1$) objects. To see if the star-like  sources are classified appropriately, we over-plotted (in cyan) the sources having high stellarity (i.e., star-galaxy classifier; sgc$>0.85$ ), measured with the \texttt{SExtractor} on the CFHT data \citep{H07,J10}.  The stellar sequence appears prominently in the optical colour space because the stellar SED shows blackbody-like behaviors, which naturally generate a systematically and statistically well-defined sequence. 

In our spectroscopic sample (as explained in sec. 3.2), we  have various types classified by the line emission diagnostics. 
We also plotted some of them here: spectroscopically confirmed stars (marked by yellow star), type-1 AGNs (red cross), AGNs identified in X-ray data (green square), as well as the sub-mm galaxies detected in SCUBA-2 survey (salmon diamond).

Figure 10 shows the colour-colour diagrams using the photometry in the HSC and AKARI NIR bands. The violet dots form a well-distinct track, cyan dots are exactly tracing this track, and five spectrocopic stars are overlaid on them. 
Those star-like sources (as well as the point sources) seem to be the representative of the Galactic stars, but it is obvious that not all of them are stars (i.e., quite a few real galaxies, which just happen to fall on the stellar locus, are included in the vicinity of the sequence). 
This implies   that the source matching was properly achieved  and the star-galaxy separation was also effectively done.   
In the optical colour-colour diagrams, the stellar sequence overlaps with extragalactic sources, positionally entangled with them on the same area (Figure 10a and 10b), but when the NIR bands are involved, this stellar sequence gets separated from the extra-galactic populations (Figure 10c, 10d). In the NIR colour-colour diagrams, however, stars seem to stay together in a rather circular/elliptical area,  not in the form of a track (Figure 11). They  gradually disappear in the longer wavelengths (Figure 12). 

In the optical colour-colour diagrams (Figure 10), the black and grey dots (local and high-z galaxies, respectively) are gathered mostly in the same area, but the grey dots seem to be more widespread.  In Figure 11, the black dots seem to be gathering in an apparently different place from the grey dots which are spread towards the redder direction, being consistent with the photo-z classification and implying high-z populations. This separation becomes obvious in $N2-N4$ colour (Figure 12b and 12c), which appears to be a very good selector of high-z objects or AGNs, seemingly related to hot dust heated by energetic sources \citep{Fritz06,Lee09}.

In Figure 11  and 12, we present the redshift tracks ($0<$z$<5.5$)  to compare with the colour-colour diagrams   based on the HSC and AKARI NIR bands (See Figure 11a$^{\prime}$,  11b$^{\prime}$, 12a$^{\prime}$, and 12b$^{\prime}$). They enable us to seize the characteristics of the sources by comparing the trajectories of typical models with real galaxies observed by the optical-IR surveys (as well as the symbols in the top panels, Figure 11a, 11b, 12a, and 12b).  In Figure 12, it should be noted that the model tracks show some overlaps in a certain area suggesting there seems to be partly a mixture of SFGs and AGNs. To select AGN types, the $N2-N4$ colour seems to be more effective when we use a combination with MIR bands (e.g., $N4-S7$ or $S7-S11$).

The AGN types and SMM galaxies stay close to the black/grey dots. It is not easy to discern clear trends. However, the green boxes are widely overlapped with all the extragalacic sources (black and grey dots), while the red crosses (type-1)  tend to stay in a specific area, all through the colour-colour diagrams. Salmon diamonds (SMM galaxies) are also spread over the black and grey dots, but more widely spread compared to the green boxes and they seem to prefer to stay around the grey dots, implying the SMM galaxies are more likely to be high-z populations.  
On the other hand, it would be interesting to see the follow-up studies (e.g., Poliszczuk et al. in prep; Chen et al. in prep) 
if   machine learning algorithms such as the support vector machine or deep neural network, etc. can do more effective separations in various color/parameter spaces (not just in two-dimensional projections of them).

\section{Summary and Conclusion }

The NEP field has been a long-standing target since it was surveyed by the legacy program of AKARI space telescope \citep{Serjeant12}.  Previous optical surveys \citep{H07,J10} were incomplete, which was a strong movitation to obtain deep Subaru/HSC optical data covering the  entire NEP field \citep{Goto17}. 
We achieved the faint detection limits \citep{Oi20}, which enabled us to identify faint AKARI sources in the near and mid-IR bands, and initiated a variety of new studies.  
We constructed a band-merged catalogue containing photometric information for 42 bands from UV to the sub-mm 500$\mu$m.  
The photo-zs for the NEPW sources were derived based on this data with all available redshift information \citep{Ho20}, and were incorporated into the catalogue as well. 
We investigated the photometric properties of the NEPW sources observed by the HSC using colour-colour diagrams based on this band-merged catalogue. 
We are able to roughly see how the shape of stellar sequence changes and which areas the AGN types prefer to stay in different colour spaces, as the observed wavelength increases, although it is difficult to tell about the clear trend of extragalactic populations because none of the quantitative analysis has been made. 

This band-merging gives us the benefits of constructing full SEDs for abundant dusty galaxy samples for SED modeling, e.g., using CIGALE \citep{Boquien19} or MAGPHYS \citep{daCunha08},  especially taking advantage of the uniqueness of the continuous MIR coverage as well as a wide range of panchromatic photometry. It provides more opportunities to disentangle otherwise degenerate properties of galaxies or to excavate hidden information for a better understanding of the physics behind various IR features.

Due to the uniqueness of the filter coverage by AKARI, this legacy data remains the only survey having continuous mid-IR imaging  until  JWST carries out its first look at the   sky. The science on this NEP field is currently driven by Subaru/HSC \citep{Oi20}, SCUBA-2 \citep{Shim20}, and homogeneous spectroscopic surveys. 
Since many future space missions are planning to conduct deep  observations of this area --  e.g.,  Euclid \citep{Laureijs11}, JWST \citep{Gardner06}, SPHEREx \citep{Dore16,Dore18},  etc.,
a great deal of synergy is expected together with the legacy data as well as our ongoing campaigns. 


\section*{Acknowledgements}
We thank the referees for the careful reading and constructive suggestions to improve this paper. 
This work is based on observations with AKARI, a JAXA project with the participation of ESA, universities, and companies in Japan, Korea, the UK. This work is based on observations made with the Spitzer Space Telescope, which is operated by the Jet Propulsion Laboratory, California Institute of Technology under a contract with NASA. Support for this work was provided by NASA
through an award issued by JPL/Caltech. Herschel is an ESA space observatory with science instruments provided by European-led Principal Investigator
consortia and with important participation from NASA.  TG  acknowledges  the  supports  by  the  Ministry  of  Science and Technology  of Taiwan through grants 105-2112-M-007-003-MY3 and 108-2628-M-007-004-MY3. HShim acknowledges the support from the National Research Foundation of Korea grant No. 2018R1C1B6008498. TH is supported by the Centre for Informatics and Computation in Astronomy (CICA) at National Tsing Hua University (NTHU) through a grant from the Ministry of Education of the Republic of China (Taiwan).

\section*{Data availability}

The band-merged catalogue in this work is available at Zenodo (https://zenodo.org/record/4007668$\#$.X5aG8XX7SuQ).  Other data addressed in this work  will be shared on reasonable request to the corresponding author.








\bsp	
\label{lastpage}
\end{document}